\documentclass[11pt, letterpaper]{article}

\usepackage[left=1in, right=1in, top=1in, bottom=1in]{geometry}
\usepackage[utf8]{inputenc}
\usepackage[T1]{fontenc}
\usepackage{bm}
\usepackage{type1cm}
\usepackage{lettrine}
\usepackage{amsmath,amssymb,amsthm}

\usepackage{moreverb}
\usepackage{mathtools}
\usepackage{amsmath}
\usepackage{amssymb}
\usepackage{algorithmic}
\usepackage{graphics}
\usepackage{graphicx}
\usepackage{caption}
\usepackage{extarrows}
\usepackage{color}
\usepackage{framed}
\usepackage{wrapfig}
\usepackage{bm}
\usepackage{mathrsfs}
\usepackage{mathabx}
\usepackage{multirow}
\usepackage{longtable}
\usepackage{hyperref}
\usepackage{paralist}
\usepackage{indentfirst}
\usepackage{relsize}
\usepackage{extarrows}
\usepackage{textcomp}
\usepackage{makecell}
\usepackage{upgreek}
\usepackage{bm}
\usepackage{mwe}
\usepackage[dvipsnames,table]{xcolor}
\usepackage{booktabs}
\usepackage{authblk}
\usepackage{lettrine}
\usepackage{type1cm}
\usepackage{threeparttable}
\usepackage[sort&compress,numbers]{natbib}
\usepackage[figurename=Figure]{caption}
\usepackage[normalem]{ulem}
\usepackage{subcaption} 
\usepackage{float} 
\usepackage{tabularx}
\usepackage{booktabs}
\usepackage{siunitx}

\graphicspath{ {./Figure/} }
\usepackage[font=footnotesize,labelfont=bf]{caption}
\newcommand{\eref}[1]{(\ref{#1})}
\providecommand{\keywords}[1]{\textbf{\textit{Keywords: }} #1}
\newcommand{\rev}[1]{{\color{black} #1}}
\newcommand{\revnew}[1]{{\color{black} #1}}

\newcommand{\revdata}[1]{{\color{black} #1}}
\newcommand{\yledit}[1]{{\color{black} #1}}

\newcommand{\revdatanew}[1]{{\color{black} #1}}

\hypersetup{
bookmarks=true,
bookmarksopen=true,
bookmarksnumbered=true,
unicode=false,
pdftoolbar=true,
pdfmenubar=true,
pdffitwindow=false,
pdfstartview={FitH},
pdftitle={My title},
pdfauthor={Author},
pdfsubject={Subject},
pdfcreator={Creator},
pdfproducer={Producer},
pdfkeywords={keywords},
pdfnewwindow=true,
colorlinks=true,
linkcolor=blue,
citecolor=blue,
filecolor=blue,
urlcolor=blue
}

\begin{document}
\renewcommand{\qedsymbol}{}
\newenvironment{breakablealgorithm}
  {
   \begin{center}
     \refstepcounter{algorithm}
     \hrule height.8pt depth0pt \kern2pt
     \renewcommand{\caption}[2][\relax]{
       {\raggedright\textbf{\ALG@name~\thealgorithm} ##2\par}%
       \ifx\relax##1\relax 
         \addcontentsline{loa}{algorithm}{\protect\numberline{\thealgorithm}##2}%
       \else 
         \addcontentsline{loa}{algorithm}{\protect\numberline{\thealgorithm}##1}%
       \fi
       \kern2pt\hrule\kern2pt
     }
  }{
     \kern2pt\hrule\relax
   \end{center}
  }
\makeatother

\title{\textbf{Lagrangian-Eulerian learning of flow field and trajectories \yledit{with TrajectoryFlowNet}}\vspace{12pt}}

\author[1]{Jingdi Wan}
\author[2]{Hongping Wang}
\author[1]{Bo Liu}
\author[2]{Xiaolei Yang}
\author[3]{Xiaodong Hu}
\author[4]{Shengze Cai}
\author[1,2]{\\Guowei He}
\author[1,2$^*$]{Yang Liu\vspace{12pt}}

\affil[1]{\small School of Engineering Science, University of Chinese Academy of Sciences, Beijing, China} 
\affil[2]{\small The State Key Laboratory of Nonlinear Mechanics, Institute of Mechanics, Chinese Academy of Sciences, Beijing, China}
\affil[3]{\small China University of Petroleum-Beijing, Beijing, China}
\affil[4]{\small College of Control Science and Engineering, Zhejiang University, Hangzhou, China \vspace{24pt}}

\affil[*]{Corresponding author: \url{liuyang22@ucas.ac.cn}}

\date{}

\maketitle

\normalsize

\vspace{-12pt} 
\begin{abstract}
\small
Predicting particle transport in complex flows is traditionally achieved by solving the Navier-Stokes equations. Numerical and experimental methods typically require deep physical insights and incur high computational costs. While machine learning offers an alternative, existing approaches fail to generalize the prediction of long-range particle trajectories in sparse data regimes. To this end, we propose TrajectoryFlowNet, a Lagrangian–Eulerian physics-informed neural network architecture, for fluid flow velocimetry and imaging via learning to predict spatiotemporal flow fields and long-range particle trajectories. The salient features of our model include its ability to handle complex flow patterns with irregular boundaries, predict the full-field flows, image the long-range flow trajectory of any arbitrary particle, and ensure physical consistency in predictions based only on very scarce measurement of flow trajectories. We validate TrajectoryFlowNet via both numerical examples (e.g., lid-driven cavity flow and complex cylinder flow) and experimental test cases (e.g., aortic and ventricle blood flows) across diverse flow scenarios. The results demonstrate our model's effectiveness in capturing intricate particle-laden flow dynamics, enabling long-range tracking of particles and accurate construction of flow fields in real-world applications.

\end{abstract}

\vspace{12pt} 

\keywords{Lagrangian–Eulerian learning, flow velocimetry and imaging, particle trajectory tracking, flow field prediction}

\vspace{12pt} 

\section*{Introduction}
Tracking and quantifying complex fluid flows plays an important role in nature and engineering applications, e.g. physics, oceanography, ecology, and pharmacokinetics~\cite{brandt2022particle, laukert2025dynamic,ser2023lagrangian,tibbittYaoWuZaiTiNeiYunShuGuoChengEmergingFrontiersDrug2016}. In such a process, the flow state is tracked by small tracer particles that are passively transported~\cite{schroder20233d}. \textcolor{black}{By assuming that} the particle behaves as an infinitesimal fluid element, the flow properties can be better understood, analyzed and predicted.
Experimental techniques like Particle Image Velocimetry (PIV), Particle tracking velocimetry (PTV), and drifters can effectively capture flow information for small-scale and mesoscale fluid dynamics~\cite{PIVgrant1997particle,PIVwesterweel2013particle,ohmi2000particle, ma2025limited}. PIV measures velocity distributions by analyzing tracer particle displacement between consecutive frames but requires high-quality, uniformly dense seeding, limiting applicability in low-density flows or specialized gases~\cite{PIVadrian2005twenty}. Moreover, PIV struggles to capture density distributions, impeding pressure field reconstruction~\cite{raffelParticleImageVelocimetry2007}, and does not provide explicit Lagrangian particle trajectories.
In contrast, PTV adopts a Lagrangian description and resolves individual particle trajectories but typically focuses on limited detectable particles and challenges reconstructing complete spatiotemporal evolution of all passive tracers~\cite{malik1993particle}. 
It is worth noting that Lagrangian particle tracking (LPT) methods like shake-the-box, enables to reconstruct trajectories for a large fraction of tracer particles~\cite{kim2023experimental} but are limited to small inter-frame displacements and struggle with partial data loss or rapid flow variation~\cite{schanz2016shake}. 
A critical challenge in turbulence theory is establishing the connection between Eulerian and Lagrangian space-time correlations, which characterizes how small-scale eddies decorrelate from both perspectives and requires solving a specific closure problem~\cite{he2017space, lumley1962mathematical}.

Recent advancements in positioning and data transmission have enhanced drifter measurements of mesoscale vortices, but spatial coverage remains inadequate with sparse data~\cite{hazaBuNengQiuZhengGeKongJianYuWindBasedEstimations2019}. In biomedicine, nanoparticle-based in vivo imaging serves as passive transport carriers for personalized medicine, yet current techniques suffer from insufficient resolution and inability to capture particle trajectories, limiting disease site analysis~\cite{ryvolova2012modern,habeeb2024enhancing}.

Numerical approaches characterize flow states by tracking passive particles through discrete partial differential equations (PDEs)~\cite{m.kuertenPointParticleDNSParticleLaden2016}. Notably, LPT, based on hydrodynamic models quantifies flow transmission in fluids~\cite{schroder20233d,kelley_lagrangian_2013,vieira_internal_2020,ser2021lagrangian,xia2013lagrangian}, enabling comprehensive analysis of fluid motion and intricate flow phenomena~\cite{kuznetsov2000passive}. However, the number of particles tracked becomes the main obstacle. As particle numbers increase, computational resources increase significantly. Meanwhile, LPT is computationally slow in unstructured grids, limiting tracked particles and modeled behaviors~\cite{crisanti1991lagrangian,rom1990analytical,ottino1990mixing}.

The revolutionary development of artificial intelligence (AI) leads to an exceptionally powerful tool for constructing data-driven models~\cite{allen2025end,binz2025foundation,kuehl2025pathology,lakeJiQiXueXiRenZhiKeXueHumanlevelConceptLearning2015}. In fluid mechanics, extensive experimental and numerical data has enabled deep learning algorithms to solve fluid flow problems~\cite{m.kuertenPointParticleDNSParticleLaden2016,wang1993settling,callaham2021learning}. LiteFlowNet~\cite{cai2019particle} and DeepPTV~\cite{liang2021deepptv} have been proposed as PIV and PTV approaches, respectively, to estimate the motion of complex flows which carry passive particles. While these methods represent effective attempts in modeling particle-laden flows, they are fundamentally based on Eulerian descriptions and cannot explicitly resolving the trajectories of individual particles. In contrast, purely Lagrangian-based data-driven approaches have demonstrated strong performance in tracking particle motion when the fluid velocity is known analytically~\cite{wei2024predicting,han2022exploratory}. However, such methods face significant challenges when applied to turbulent flows due to the large volume of data required for accurate predictions~\cite{hassanian2023deciphering,hassanian2025data}.
Due to the over-parameterized and black-box nature of the models, issues arise in interpretability, overfitting, and generalization. Embedding knowledge of fluid mechanics to form physics-informed learning has shown potential in alleviating these fundamental issues, leading to more reliable predictions.

Physics-Informed Neural Networks (PINNs) represent notable progress for forward and inverse analyses of partial differential equation systems~\cite{raissi2019physics}. By embedding prior physics knowledge, PINNs reduce reliance on large training data volumes, improving prediction accuracy~\cite{raissi2020hidden, karniadakisPhysicsinformedMachineLearning2021,chen2021physics}. Using particle tracking data, PINNs reconstruct high-resolution velocity fields, learn three-dimensional flow dynamics, predict turbulent boundary layers, and infer pressure fields~\cite{wang2022dense,cai2024physics,hasanuzzaman2023enhancement,wang2025machine}. PINNs fuse PIV and point-wise probe data for high temporal resolution in particle-laden flow estimation~\cite{soto2024complete,lai2024temporal}, with derivative methods like AI velocimetry demonstrating strong capability in modeling realistic particle-carrying flows~\cite{cai2021artificial,boster2023artificial,toscano2024inferring}. \rev{Besides, Artificial Intelligence Velocimetry-Thermometry (AIVT)~\cite{toscano2025aivt} has been developed to simultaneously reconstruct continuous velocity and temperature fields in turbulent thermal convection. By integrating physics-informed Kolmogorov-Arnold networks with residual-based adaptive resampling, AIVT enables accurate inference of velocity and temperature fields (along with their gradients) from sparse Lagrangian particle trajectories, without needing direct temperature measurements in training. Importantly, AIVT still relies on LPT measurements to obtain particle trajectories. }

Although these methods produce flow fields consistent with PIV or PTV observations, they lack explicit treatment of Lagrangian particle trajectories. Recently, graph neural networks like GotFlow3D~\cite{liang2023recurrent} have been employed to learn flow motions using optimal transport graphs for structured representations, estimating transient particle trajectories from adjacent frames while inferring flow characteristics. \rev{However, such a method is a double-frame motion estimator designed to predict instantaneous particle displacements between two adjacent time points. Because it only correlates image intensities over a pair of consecutive frames and does not identify or follow individual particles across multiple frames, PIV cannot establish persistent particle identities over time and is therefore unsuitable for long-range trajectory prediction.}

To this end, we propose TrajectoryFlowNet, a Lagrangian–Eulerian physics-informed neural network architecture, for fluid flow velocimetry and imaging via learning to predict the spatiotemporal flow field and long-range trajectory of any arbitrary particle. Specifically, TrajectoryFlowNet integrates several key innovations: (1) a hybrid learning architecture that combines Lagrangian and Eulerian descriptions, thereby enabling reconstruction of both individual particle trajectories and continuum flow dynamics; (2) maintaining physical consistency in predictions regularized by a set of prior physics; and (3) Fourier feature mapping that overcomes spectral bias in neural networks, enabling accurate learning of high-frequency components and enhancing the capability to model complex, unsteady, and highly viscous flows such as blood circulation. Our model can effectively handle complex flow patterns with irregular boundaries, predict the full field of flows, and reconstruct trajectories of arbitrary particles based on sparsely observed trajectory data. Through extensive experiments on a variety of numerical and experimental examples, we demonstrate the effectiveness of TrajectoryFlowNet in simultaneously tracking particles and predicting flow fields, making it a promising tool for sensing and data assimilation in the field of fluid dynamics.

\section*{Results}

\subsection*{The overall framework of TrajectoryFlowNet}

\begin{figure}[h!]
 \centering
  \includegraphics[width=1.0\linewidth]{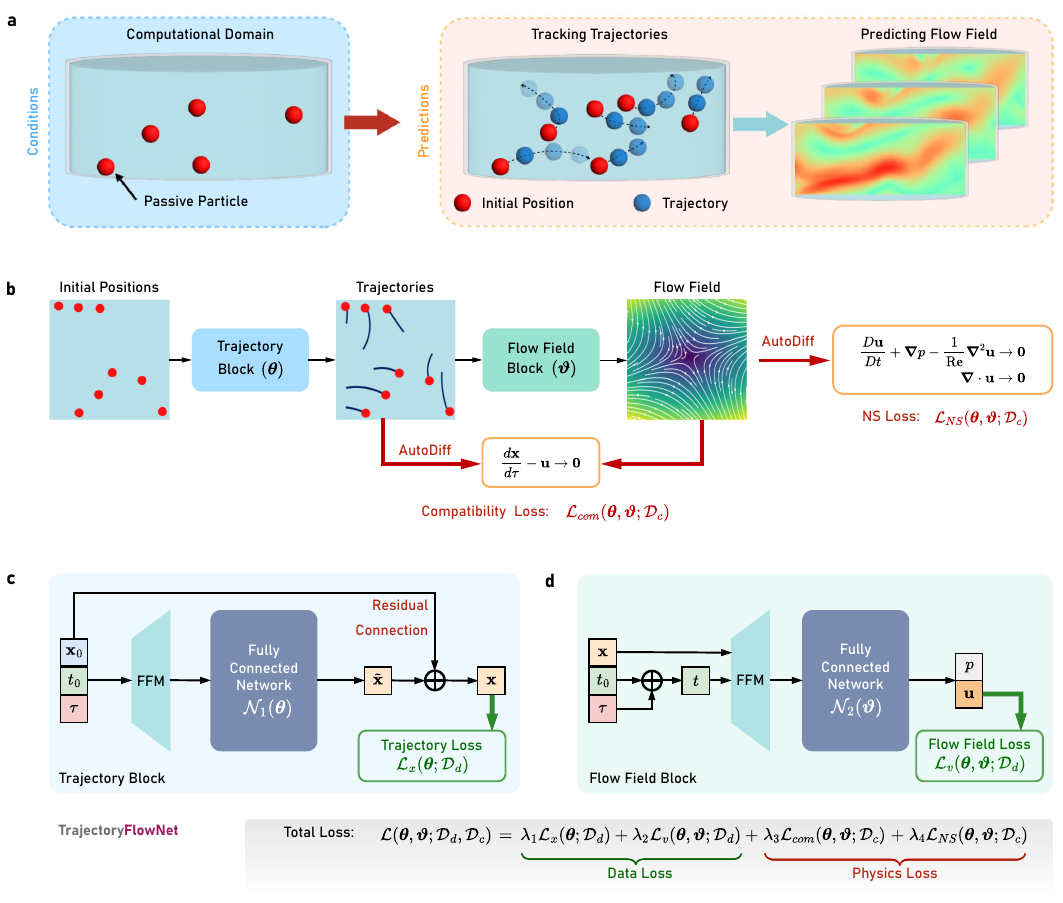}
   \caption{
   \textbf{Overview of TrajectoryFlowNet. a}, Schematic of simultaneous prediction of particle trajectories and flow field. The computational domain contains only the initial positions of passive particles (red dots). The objective is to predict the trajectories of these particles (blue dots) and infer the spatiotemporal flow field. \textbf{b}, Schematic architecture of TrajectoryFlowNet, which consists of a Trajectory Block and a Flow Field Block. \textbf{c}, Schematic of the Trajectory Block (with trainable parameters $\boldsymbol{\theta}$), which maps the initial position, release time and motion time $\{\mathbf{x}_0,t_0,\tau\}$ of a particle to its displacement $\tilde{\mathbf{x}} = \tilde{\mathbf{x}}(\mathbf{x}_0, t_0, \tau)$ after Fourier feature mapping. By adding this displacement to the initial position, we obtain the particle's trajectory $\mathbf{x} = \tilde{\mathbf{x}} + \mathbf{x}_0$. \textbf{d}, Schematic of the Flow Field Block, controlled by trainable parameters $\boldsymbol{\vartheta}$, which takes the Fourier feature-mapped particle trajectories and motion times (predicted by Trajectory Block) as input and learns the mapping to the flow velocity $\mathbf{u}$ and pressure $p$. Using automatic differentiation (AD), we compute the derivative of the particle position $\mathbf{x}$ with respect to the particle motion time $\tau$ (aka, $\mathrm{d}\mathbf{x}/\mathrm{d} \tau$) at machine precision. The residual between this derivative and the network-predicted velocity is used to enhance physical consistency of the network. Note that the input-output variables are nondimensionlized. The predicted flow velocity and pressure fields are assumed to satisfy the Navier-Stokes (NS) equations, leading to the NS loss. The total loss function includes data losses $\mathcal{L}_x$ (trajectory loss) and $\mathcal{L}_v$ (velocity loss), as well as the residual physics losses $\mathcal{L}_{com}$ (compatibility loss) and $\mathcal{L}_{NS}$ (NS equation loss). Here, $\mathcal{D}_d$ represents the training data, and $\mathcal{D}_c$ the collocation points. \revdatanew{$\lambda_1$,  $\lambda_2$, $\lambda_3$, $\lambda_4$ denote the loss hyperparameters.}
}
   \label{fig:1}
\end{figure}

We examine an infinitesimal fluid parcel composed of a microscopic lightweight particle with negligible mass and size, assuming that such a particle instantaneously adapts to any changes in the fluid velocity. Our objective is to develop a machine learning model to simultaneously predict the evolving velocity and pressure fields as well as the long-range trajectory of any arbitrary particle (see Fig. \ref{fig:1}\textbf{a}). To this end, we propose TrajectoryFlowNet (see Fig. \ref{fig:1}\textbf{b}--\textbf{d}), a Lagrangian–Eulerian deep learning framework that jointly models particle trajectories and fluid flow velocimetry, enabling robust prediction under scarce training data conditions.

We hypothesize that the state of each particle is uniquely identified by its initial release position ($\mathbf{x}_0\in \mathbb{R}^n$) and time ($t_0\in \mathbb{R}$), where $n$ denotes the spatial dimension. Hence, the trajectory of a moving particle at Lagrangian time $\tau\in \mathbb{R}$ (elapsed since the release of the particle, $\tau := t - t_0$) can be expressed as $\mathbf{x}(\mathbf{x}_0,t_0,\tau) = \mathbf{x}_0 + g(\mathbf{x}_0,t_0,\tau)\in \mathbb{R}^n$, where $g$ denotes a mapping function and $t\in \mathbb{R}$ the Eulerian time. The velocity of the parcel carrying the particle, namely, $\mathbf{u}= d \mathbf{x} / d\tau \in \mathbb{R}^n$, naturally bridges the Lagrangian and Eulerian viewpoints. The dynamics of these parcels follow the Navier-Stokes equation (NSE) in Eulerian form, shown in Eq. (\ref{NSeq}).

The schematic architecture of TrajectoryFlowNet is depicted in Fig. \ref{fig:1}\textbf{b}, which comprises two core components: the Trajectory Block and the Flow Field Block. The Trajectory Block shown in Fig. \ref{fig:1}\textbf{c} predicts particle displacements from their initial positions, while the Flow Field Block illustrated in Fig. \ref{fig:1}\textbf{d} infers the corresponding spatiotemporal velocity and pressure fields based on the predicted trajectories. Notably, we incorporate Fourier feature mapping at each block's input to enhance the model's capability of resolving high-frequency components in the flow.

To ensure physical consistency of the model, we leverage automatic differentiation to enforce physics constraints. The derivative of particle position with respect to the Lagrangian time is matched to the predicted velocity (aka, $d \mathbf{x} / d\tau - \mathbf{u} \rightarrow \mathbf{0}$), while the predicted flow fields ($\mathbf{u}$ and $p$) are set to satisfy he the Navier-Stokes equation as shown in Eq. \eref{NSeq}. Since the measurement data inherently includes specific initial and boundary conditions (I/BCs), these constraints, regardless of \textit{a priori} known or empirically obtainable, do not need to be explicitly defined during our model construction. More details on the theoretical background, the network architecture, and the loss functions for model training can be found in the \textcolor{blue}{Methods} section.

To demonstrate the efficacy of TrajectoryFlowNet in simultaneously tracking long-range trajectories of passive particles and inferring flow field characteristics, we have considered four distinct systems: the lid-driven cavity flow, the cylinder flow in a container with complex geometry, the aortic blood flow, and the left ventricle blood flow. The training datasets for the first two systems were obtained by spatiotemporal down-sampling of high-fidelity simulations that encompass various I/BCs, number of passive particles, Reynolds numbers, and flow regimes (see \textcolor{blue}{Supplementary Note C.1} and \textcolor{blue}{C.2}). In contrast, the training datasets for the rest two systems were collected from real experiments as detailed in \textcolor{blue}{Supplementary Note C.3} and \textcolor{blue}{C.4}. We also evaluated the impact of key architectural and encoding choices on model performance. Specifically, we compared the use of Fourier feature mapping (see \textcolor{blue}{Supplementary Note D.1}), the effect of separate versus joint training strategies (see \textcolor{blue}{Supplementary Notes D.2}), and the influence of different activation functions (see \textcolor{blue}{Supplementary Note D.3}). 

\subsection*{Lid-driven cavity flow}

Square cavity flow has been a classical problem in fluid mechanics, which has a simple geometric structure and includes various fluid dynamics phenomena, such as boundary layer separation, cavity vortices, and self-sustained oscillations. Hence, we take the lid-driven cavity flow as a  benchmark to verify the effectiveness of our method in simultaneously predicting the long-range trajectories of particles and the velocity-pressure fields.

\begin{figure}[h!]
 \centering
  \includegraphics[width=1.0\linewidth]{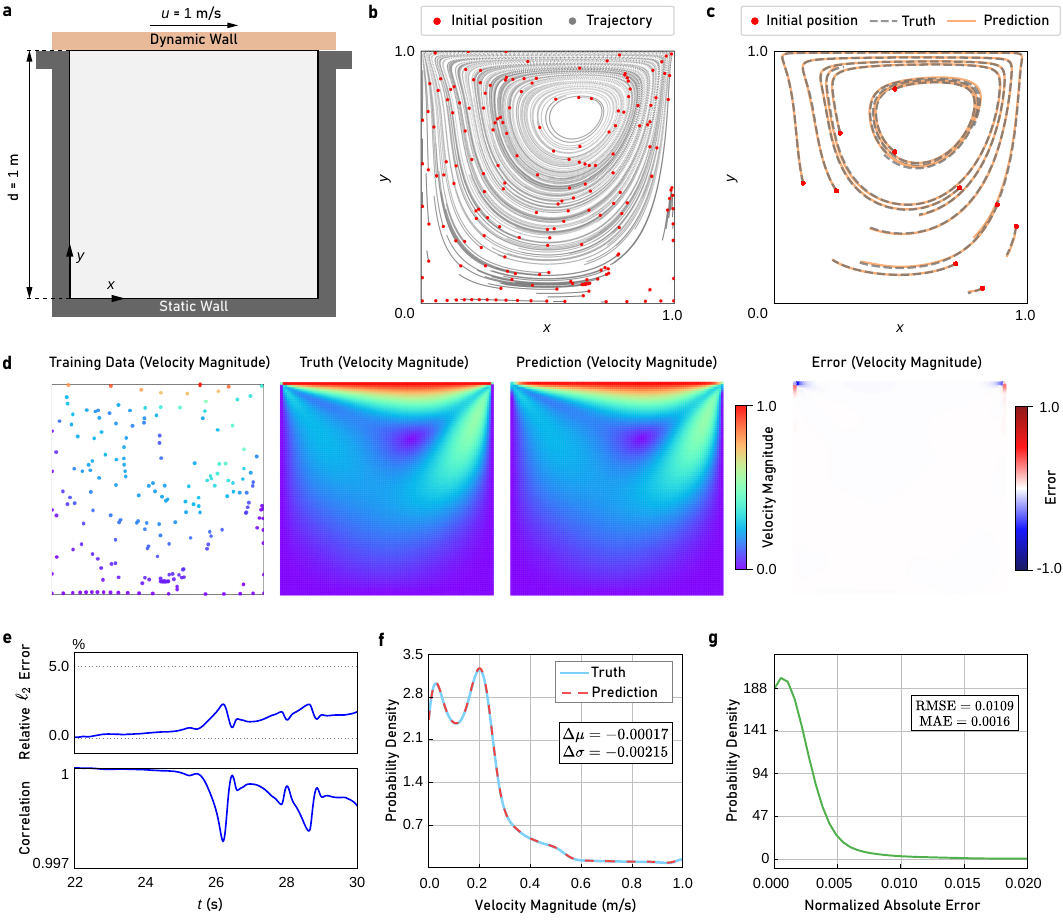}
   \caption{
   \textbf{Results of the Lid-driven Cavity Flow Case.} 
   \textbf{a}, Description of lid-driven cavity flow test setup. A $1\times1$ square cavity model was established, with fixed left, right, and bottom walls, and a moving top lid from left to right at an initial speed of 1 m/s, resulting in a flow with a Reynolds number of 100. The problem was simulated by FLUENT. At $t_0 = 5$ s, 300 particles are released within the cavity, and their trajectories, velocities, and pressure data are recorded. 
   \textbf{b}, Illustration of typical particle trajectories in the training set. Red dots denote the initial particle positions, and gray dots the trajectory points.
   \textbf{c}, Illustration of randomly selected predicted particle trajectories in the test set. The red dots represent the initial positions of the passive particles, the orange solid lines the predicted trajectories by TrajectoryFlowNet, and the dashed lines the true values.
   \textbf{d}, The flow velocity magnitude field $t = 25$ s (from left to right: the velocity magnitudes of all passive particles in the training set, snapshot of the ground truth, snapshot of the prediction by TrajectoryFlowNet, and the prediction error).
   \revdatanew{\textbf{e}, Temporal evolution of the relative $\ell_2$ error (up) and Pearson correlation coefficient (down) for predicted trajectory displacement magnitude $\left| \mathbf{x} \right|$ in the test set. Over the considered time period, the relative $\ell_2$ error remains below 5\%, while the Pearson correlation coefficient remains above 0.997, demonstrating the high accuracy and temporal consistency of the model’s predictions.}
   \textbf{f}, Distribution of the predicted velocity magnitudes for the text set at $t = 25$ s via Kernel density estimation (KDE). The solid blue line represents the ground truth, and the dashed red line the prediction by TrajectoryFlowNet. 
   \textbf{g}, KDE distribution of the predicted velocity magnitude errors at $t = 25$ s for the test set. The errors are normalized by the dynamic range of the respective velocity components as the characteristic scale.
}
   \label{fig:2}
\end{figure}

In particular, we consider a 1 m $\times$ 1 m square cavity as shown in Fig. \ref{fig:2}\textbf{a}, with fixed left, right, and bottom walls, and a moving top lid from left to right at an initial speed of 1 m/s (corresponding to a Reynolds number of 100). We used FLUENT for simulation with a uniform grid of $100\times100$. Under the influence of the moving top lid, a large vortex formed near the geometric center of the cavity, with smaller flows at the bottom and side edges. At $t_0=5$ s, 300 particles were dispersed inside the cavity, and their trajectories, as well as the velocity and pressure at their locations, were recorded. The total motion duration was set to be 30 s. The training set consists of 80,200 data points randomly selected from 200 particles within the time interval of [22, 30] s (see Fig. \ref{fig:2}\textbf{b}). The test set includes 40,100 data points from the rest 100 particles. \revnew{The training and test sets consist of non-overlapping spatiotemporal samples drawn from the same ergodic flow field, ensuring identical statistical properties across both datasets.} A set of 160,400 collocation points were generated in the spatiotemporal domain by Latin hypercube sampling to enhance the physics-constrained training. The model training was performed by L-BFGS. 

\begin{table}[h!]
\centering
{\caption{Summary of TrajectoryFlowNet's performance in terms of prediction accuracy \revnew{using two-sided Pearson correlation tests} ($x$ and $y$ denote the particle trajectory coordinates, $u$ and $v$ the flow velocity field, and $p$ the pressure field).}\label{table1}}
\small
   \begin{tabular}{cccccc} 
        \toprule
        \multirow{2}{*}{Verifiable Example}&\multicolumn{5}{c}{Pearson Correlation}\\
        \cmidrule{2-6}& $x$ & $y$ & $u$ & $v$ & $p$\\
        \midrule
        Lid-driven cavity flow&0.9987&0.9996&0.9998&0.9998&0.9954\\
        Complex cylinder flow&0.9975&0.9856&0.9814&0.9787&0.9614\\
        Experimental aortic blood flow&0.9965&0.9983&0.8703&0.7355&--\\
        Experimental left ventricle blood flow&0.9988&0.9974&0.7316&0.9201&--\\
        \bottomrule
   \end{tabular}
   \vspace{0pt}
\end{table}

The performance of TrajectoryFlowNet in terms of prediction accuracy across all four cases is summarized in Table \ref{table1}, showcasing the model's robustness and high accuracy. Fig. \ref{fig:2}\textbf{c} displays the predicted particle trajectories by TrajectoryFlowNet, while
\rev{Fig. \ref{fig:2}\textbf{d} illustrates the training data and the predicted flow velocity magnitude at $t = 25$ s. Despite with sparse and non-uniform training data, the predicted velocity field well matches the ground truth. The main errors occur near the top corners of the cavity, which may be attributed to data scarcity in these regions where less particles tend to reach these locations. 
\revdatanew{Fig. \ref{fig:2}\textbf{e} shows the corresponding relative  $\ell_2$ error and the Pearson correlation for the predicted trajectory displacement magnitude $\left| \mathbf{x} \right|$ over time across all test sequences. It can be seen that the relative  $\ell_2$ error remains consistently low throughout the trajectory evolution, staying below 5.0\%, while the correlation remains consistently above 0.997, demonstrating the stability of our method in long-range tracking of particle trajectories.}}

Fig. \ref{fig:2}\textbf{f} displays the kernel density estimate (KDE) of flow velocity magnitude field at $t = 25$ s, characterizing the statistical distribution of predicted velocities. The close alignment between the two curves, together with their sharp peaks and narrow spreads, indicates that the predicted results are closely aligned with the ground truth, further evidenced by the small absolute differences in mean \revdatanew{($|\Delta \mu|=0.00017$) and standard deviation ($|\Delta \sigma|=0.00215$)}. The normalized error distribution of the predicted velocity magnitudes at $t = 25$ s is shown in Fig. \ref{fig:2}\textbf{g}. It can be seen that the prediction errors are predominantly below 0.005, with the overall root mean square error (RMSE) and mean absolute error (MAE) values smaller than 0.011.

\rev{We also compared TrajectoryFlowNet with the AIV \cite{cai2021artificial} and AIVT~\cite{toscano2025aivt} method for flow field prediction (see Extended Data Fig. \textcolor{blue}{1} and Extended Data Table \textcolor{blue}{1}).}
\rev{For a fair comparison, TrajectoryFlowNet and the AIV method employ identical hidden-layer architectures.
Meanwhile, both TrajectoryFlowNet and AIVT are trained on the same set of particle trajectories and their associated velocity measurements. Notably, AIV and AIVT methods incorporates extra BCs during training. Furthermore, to align with our problem setting, we adapted the AIVT framework by replacing its original physical constraints with 2D incompressible Navier-Stokes equations.}

Since the pressure data are intentionally excluded from the training set, the estimated pressure field may differ from the ground truth by an additive constant. Therefore, we randomly selected a spatiotemporal point to determine the constant correction term for pressure. Although the AIV method also yielded an excellent prediction of the flow field ($u$, $v$ and $p$), the error snapshots show that TrajectoryFlowNet exhibits superior fidelity details. The accuracy of the AIV method may be influenced by factors such as the fixed number of passive particles. In regions where the flow velocity is high, particles move quickly and are distributed sparsely, making it challenging for the AIV method to achieve precise predictions. Notably, TrajectoryFlowNet can explicitly predict the long-range trajectories of passive particles, a capability not offered by the AIV method. This feature provides a new solution for particle-based flow imaging and prediction.
\rev{However, AIVT performs noticeably worse in this case. We attribute this degradation primarily to the sparsity of training data, compounded by the limited numerical accuracy of automatic differentiation in cKAN networks, which compromises the enforcement of physics-informed constraints and leads to inaccurate flow field predictions.}

\subsection*{Complex cylinder flow}

\begin{figure}[h!]
 \centering
  \includegraphics[width=0.97\linewidth]{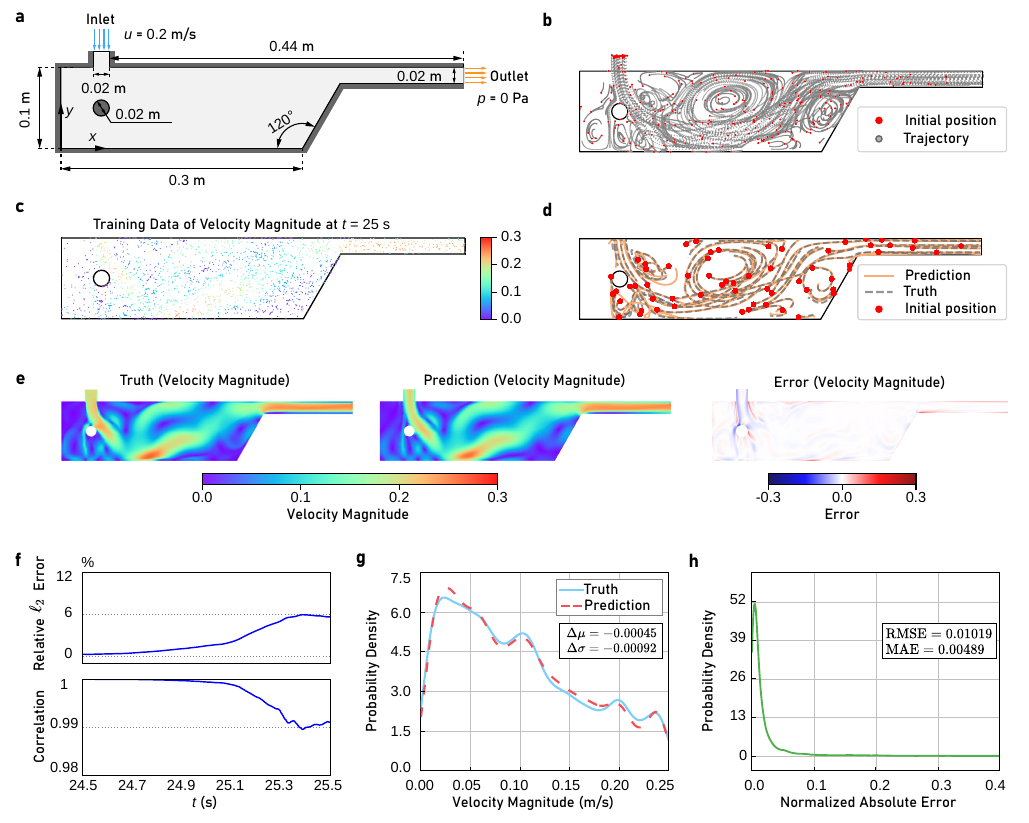}
   \caption{\rev{
   \textbf{Results of the Complex Cylinder Flow Case.}
   \textbf{a}, Description of complex cylinder flow setup. The cylinder is placed inside a duct featuring an inlet and an outlet, with liquid water as the working fluid. A steady inlet velocity of $0.2$ m/s was employed, resulting in a Reynolds number of 146,404, with zero outlet pressure. At $t=10$ s, 2200 tracer particles are released at the inlet following by tracking their positions and velocities. This case was simulated by DNS. 
   \textbf{b}, An example set of 300 particle trajectories in training set. Red dots denote initial particle positions, and gray dots the trajectory points.
   \textbf{c}, Snapshot of the flow velocity magnitudes of all particles of training set at $t = 25$ s. 
   \textbf{d}, Illustration of randomly selected predicted particle trajectories in test set. Red dots marking the initial positions, orange solid lines the predicted trajectories, and dashed lines the true values.
   \textbf{e}, The flow velocity magnitude field at $t = 25$ s (from left to right: snapshot of the ground truth, snapshot of the prediction, and the corresponding prediction error). 
   \revdatanew{\textbf{f}, Temporal evolution of the relative $\ell_2$ error (up) and Pearson correlation coefficient (down) for predicted trajectory displacement magnitude $\left| \mathbf{x} \right|$ in the test set.}
   \textbf{g}, KDE distribution of predicted velocity magnitudes in test set at $t = 25$ s. The solid blue line represents the ground truth, and the dashed red line the prediction by TrajectoryFlowNet. 
   \textbf{h}, KDE distribution of normalized absolute errors for velocity magnitudes at $t = 25$ s in the test set. 
}}
   \label{fig:3}
\end{figure}

To further examine the prediction performance of TrajectoryFlowNet, we consider a more complex flow system, namely, a 2D flow scenario in an irregular domain with a cylinder, as depicted in Fig. \ref{fig:3}\textbf{a}. The cylinder is placed inside a duct featuring an inlet and an outlet, with liquid water as the working fluid. A steady inlet velocity of $0.2$ m/s was employed, resulting in a Reynolds number of 146,404, with zero outlet pressure. \rev{The simulation data was generated by direct numerical simulations (DNS).} 

Under such conditions, the presence of the internal cylinder induces flow separation, generating complex vortices and recirculation zones near the boundaries. Passive tracer particles were released at the inlet continuously from $t = 0$ s until the end of the measurement period, with each particle’s release time serving as its initial time $t_0$. The measurement data were collected from the period of [24.5, 25.5] s. A group of 2,000 particle trajectories together with their velocity profiles were used for model training (see the illustration example of 300 particle trajectories in Fig. \ref{fig:3}\textbf{b}), while the test dataset consists of 200 randomly selected particles. Fig. \ref{fig:3}\textbf{c} displays the snapshot of the flow velocity magnitude for these 2,000 particles at $t = 25$ s. Noteworthy, the specific boundary conditions (BCs) are assumed unknown in our model training. More details of the data description can be found in \textcolor{blue}{Supplementary Note C.2}.

Fig. \ref{fig:3}\textbf{d} shows some of the predicted particle trajectories compared with the ground truth in the test dataset, 
\revdatanew{while Fig. \ref{fig:3}\textbf{f} depicts the trajectory prediction errors and Pearson correlation of the predicted displacement magnitude $\left| \mathbf{x} \right|$ for 200 test particles. Over time, the relative $\ell_2$ error remains below 6\%, with Pearson correlation coefficient staying above 0.98.}
Fig. \ref{fig:3}\textbf{e} depicts the predicted velocity magnitude field at $t = 25$ s, while Fig. \ref{fig:3}\textbf{g} illustrates the corresponding KDE and Fig. \ref{fig:3}\textbf{h} the distribution of normalized absolute errors. \rev{These results show that our model successfully predicts the flow states, with an acceptable accuracy \revdatanew{(e.g., $|\Delta \mu| = 0.00045$ and the vast majority of normalized absolute errors below 0.1)}. However, the KDE distribution of predicted velocity magnitudes shows a slight over-prediction in the low-velocity regime (velocity magnitude $<0.05$m/s) compared to the ground truth. This may be caued by the non-uniform spatial distribution of observed particles. In such regions, sparse particle trajectories provide limited supervision, making model less sensitive to the variations of flow speed.}

\rev{We again compared our TrajectoryFlowNet predictions with those by the AIV \cite{cai2021artificial} and AIVT \cite{toscano2025aivt} methods, as shown in Extended Data Fig. \textcolor{blue}{2} and Extended Data Table \textcolor{blue}{1}. It is noted that both AIV and AIVT methods requires the \textit{a priori} knowledge of BCs, while our method does not. The results show that the AIV method completely fails to predict the complex flow field, \revdatanew{yielding the Pearson correlation coefficients of 0.0586 for $u$ and 0.3335 for $v$.} In this case, the complex vortex dynamics involving generation, shedding, and dissipation, at a high Reynolds number, render reliable velocity reconstruction solely from image intensity challenging. 
In contrast, AIVT achieves relatively better velocity field predictions, benefiting from the sufficient number of observed particles in this scenario. However, due to the intrinsic limitations of cKAN networks (particularly less accuracy in automatic differentiation) it struggles to recover an accurate pressure field in regions devoid of particle data, relying solely on boundary conditions and physics constraints. This highlights a key weakness of AIVT method when observational coverage is incomplete.}

Nevertheless, TrajectoryFlowNet successfully identifies the trend of the flow field motion, demonstrating its superior capacity in both accurate prediction of long-range particle trajectories and reliable reconstruction of flow fields.

\begin{figure}[h!]
 \centering
  \includegraphics[width=0.995\linewidth]{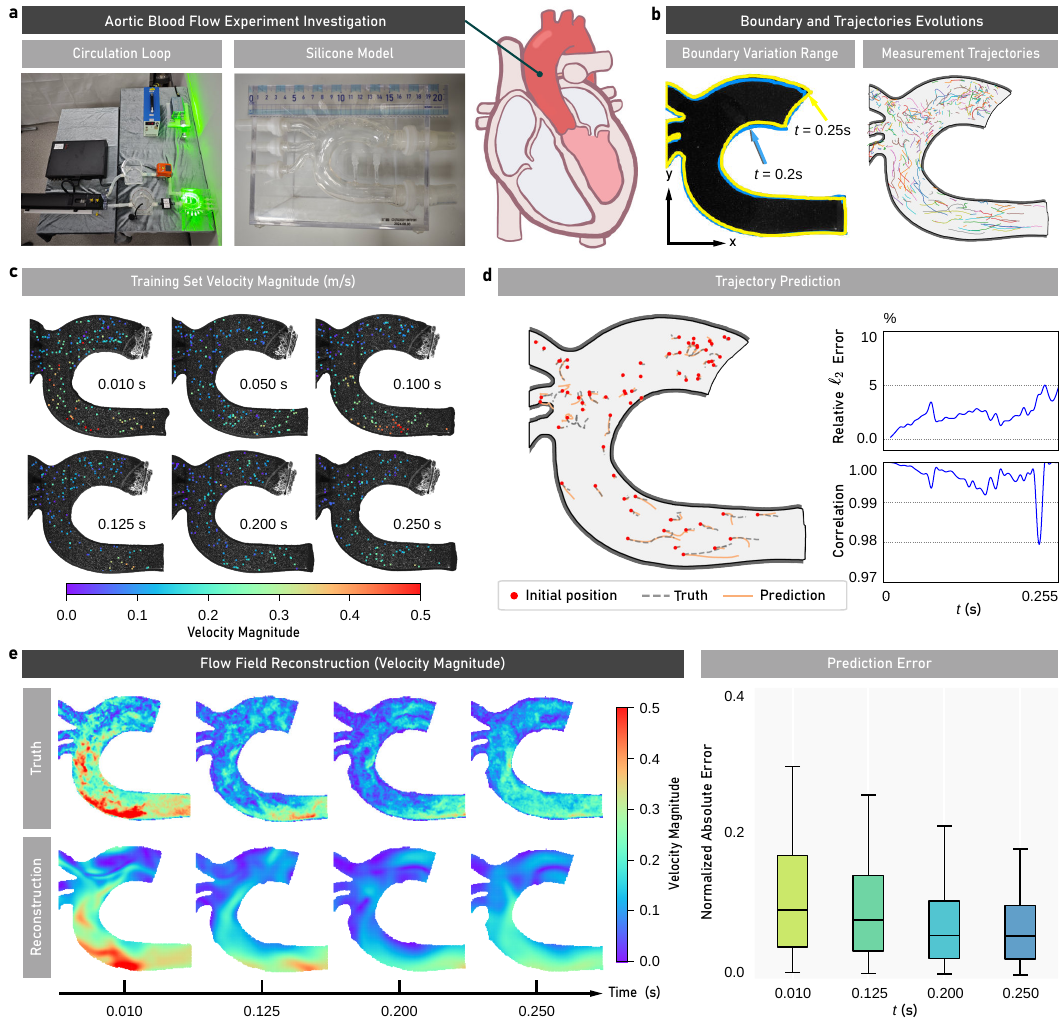}
   \caption{
   \textbf{Results of the Experimental Aortic Blood Flow Case.}
   \textbf{a}, The experiment setup, including the 3D-printed silicone model and the circulation loop.
   \textbf{b}, Illustration of the time-variant boundary of the aortic vessel and the particle trajectories in the training set.
   \textbf{c}, Snapshots of measured flow velocity magnitudes at a limited number of particles in the training set.
   \revdatanew{\textbf{d}, Predicted particle trajectories (left) along with the evolution of relative  $\ell_2$ errors and Pearson correlation coefficient of displacement magnitude $\left| \mathbf{x} \right|$ (right) for the test set.}
   \revdatanew{\textbf{e}, Snapshots of the reconstructed flow velocity magnitude fields (left) along with normalized absolute errors of predicted fields on the test set (right).} The snapshots display both the ground-truth flow field and its reconstruction. The central box spans the interquartile range from 25\% to 75\%, with the median marked by a horizontal line within the box. Lines extending from the box denote the 5\% to 95\% range. \revdatanew{ Sample sizes for the four time steps (from left to right) are $14,064$, $13,883$, $13,582$, and $13,372$ grid points.}
}
   \label{fig:4}
\end{figure}

\subsection*{Experimental aortic blood flow}
Despite the satisfactory performance of TrajectoryFlowNet on idealized simulations, testing its capability to handle real-world collected data is essential, especially under uncertainties caused by measurement noise, varying BCs, and source conditions. For instance, the hemodynamic characteristics within the aorta are expected to be influenced by aortic displacement due to the periodic beating of the heart and its dynamic changes during systole. Therefore, we herein consider a real-world problem in the biomedical field, namely, predicting the flow dynamics in the aorta for balloon-expandable transcatheter aortic valve replacement (TAVR) (see Fig. \ref{fig:4}\textbf{a}). 

\rev{TAVR is a minimally invasive procedure used to treat severe aortic stenosis, a common and life-threatening heart valve disease~\cite{cribier2016development,makkar2012transcatheter,reardon2017surgical}. However, the hemodynamics in the aortic root are highly complex due to its intricate anatomy and the combined effects of cardiac motion, vessel elasticity, and valve calcification~\cite{loukas2014anatomy}. Accurately assessing hemodynamic responses to optimize valve selection, implantation position, and deployment strategy, as well as to reduce complications, has become an urgent clinical need~\cite{khodaei2023early,garber2023impact}.
}

The dataset was collected in the lab based on a silicone model of the aorta 3D printed from CT scans of a healthy individual. This model was then submerged in a working fluid composed of 62\% water and 38\% glycerol by volume, with 230 g/L urea added by weight to match the refractive index~\cite{ChenandDasi2023}. A 23 mm Edwards SAPIEN transcatheter aortic valve (TAV) was deployed at the annular position of the aortic root model. During systole, the blood-mimicking fluid flowed from the left ventricular compliance chamber through the TAV towards the aortic root, and then back to the left ventricle via resistance valves, a fluid reservoir, and a mitral valve. Time-resolved 2D PIV experiments were conducted to measure the flow downstream of the TAV along the centerline of the root. The pairs of particle images were processed to estimate the velocity field, while applying Gaussian filtering to reduce random noise. The long-range particle trajectories were manually tracked. More details about the experimental dataset can be found in \textcolor{blue}{Supplementary Note C.3}.

It is worth noting that the irregular moving boundaries of the aortic model pose a significant challenge for traditional physics-based models to replicate realistic conditions. For example, Fig. \ref{fig:4}\textbf{b} shows the range of variation in the model boundaries. This issue can be naturally handled by point-wise neural networks. The experiment recorded the motion states of 5,489 particles, consisting of 5,117 single‑time‑step particles measured by PIV and 372 manually annotated particles tracked over multiple ($\geq 2$) time steps (see Fig. \ref{fig:4}\textbf{b}).  
To maximize the utility of the training data, we partitioned the trajectories into shorter overlapping segments spanning 2--5 time steps. \revdatanew{This augmentation was conducted solely based on the original training trajectories without introducing any additional data, expanding the dataset by 8,908 samples to a final count of 47,590 spatiotemporal points.}
The snapshots of the sparse velocity fields are depicted in Fig. \ref{fig:4}\textbf{c}. Due to the inability of sparse particle observations to fully characterize the flow field properties, we constructed two separate test sets, one for trajectory prediction and the other for velocity field reconstruction. \revdatanew{The trajectory test set consists of 69 manually annotated particle trajectories (totaling 582 data points), while the velocity test set comprises 50 PIV snapshots (spanning 0.25 s) containing all data points not used in training.}
The model was trained by L-BFGS with the tolerance gradient of $1\times 10 ^{-5}$. 

Fig. \ref{fig:4}\textbf{d} shows the predicted particle trajectories along with the corresponding relative $\ell_2$ error and Pearson correlation for the test set. \revdatanew{The maximum relative $\ell_2$ error of displacement magnitude  ($\left| \mathbf{x} \right|$) remains below 6\%, while the Pearson correlation coefficient exceeds 0.99 throughout most of the trajectory evolution.}
\revdatanew{Fig. \ref{fig:4}\textbf{e} depicts several typical snapshots of reconstructed velocity magnitude fields alongside normalized prediction errors evaluated on the test set.} It can be seen that both the multi-step trajectories and the evolving aortic flow patterns are well predicted by our model. Even in the presence of unknown uncertainties, \revdatanew{as shown in Table \ref{table1}, the Pearson correlation value of the predicted passive particle trajectories exceeds 0.99, while the predicted velocities yield a value above 0.73.} Without imposing explicit BCs, our model demonstrates the capability to handle irregular moving boundaries. Although this real-world example contains measurement noise and other numerous elusive variables not present in our training data (e.g., vessel wall elasticity and friction, blood aggregation, compressible blood density), our model demonstrated satisfactory ability to handle such uncertainties. 
\rev{Since the pressure field cannot be measured experimentally, the ground-truth pressure distribution is unavailable. To quantify the uncertainty of pressure predictions, we employed a physics-constrained Bayesian training strategy~\cite{sun2020physics}, which incorporates the 2D incompressible Navier-Stokes equations as physical constraints in the likelihood function and leverages a Stein variational gradient descent (SVGD) scheme to approximate the posterior distribution. The results of probabilistic prediction are shown in Extended Data Fig. \textcolor{blue}{3}\textbf{a}. It can be seen that the deviations of pressure prediction remain at a reasonably low level.}

Besides the aforementioned uncertainties, the prediction errors might also stem from other sources. For example, modeling the blood flow as incompressible fluid introduces approximation errors, as physiological flows exhibit subtle compressibility and non-Newtonian characteristics. Moreover, the Reynolds number varies spatially across the domain (see \textcolor{blue}{Supplementary Fig. S.2}), reflecting the underlying flow complexity. While our framework successfully identifies the Reynolds number as a relevant parameter, we identified it as a constant value for practical implementation, resulting in modeling errors. Furthermore, while the experimental model is 3D, the measurement data are derived from 2D projections. Consequently, simplifying this problem as 2D leads to inevitable modeling errors. Despite these simplifications and measurement uncertainties, the prediction results show statistically consistent agreement with the ground-truth measurement, capturing the intrinsic flow dynamics.

\begin{figure}[h!]
 \centering
  \includegraphics[width=0.995\linewidth]{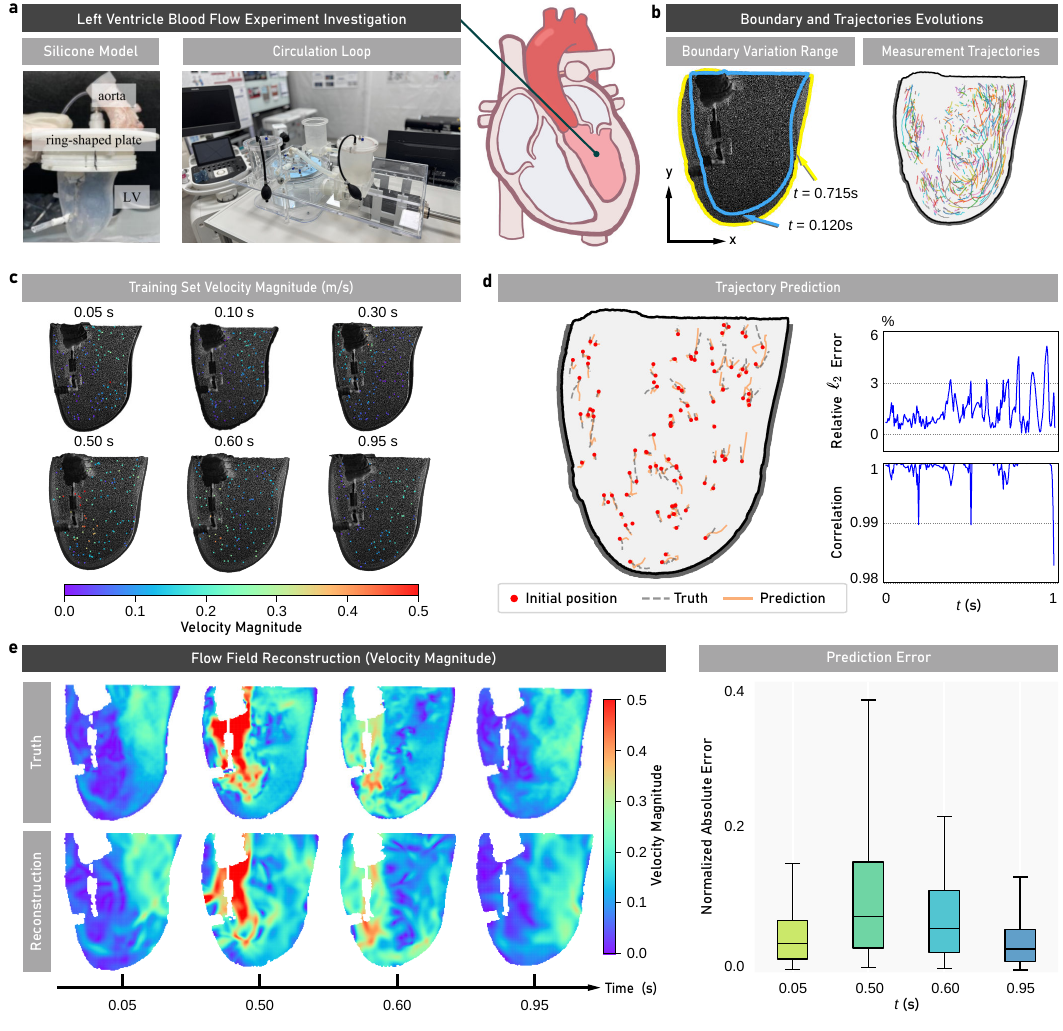}
   \caption{
   \textbf{Results of the Experimental Left Ventricle Blood Flow Case.}
   \textbf{a}, The experiment setup, including the 3D-printed silicone model and the circulation loop.
   \textbf{b}, Illustration of the time-variant boundary of the left ventricle and the particle trajectories in the training set.
   \textbf{c}, Snapshots of measured flow velocity magnitudes at a limited number of particles in the training set.
   \revdatanew{\textbf{d}, Predicted particle trajectories (left) along with the evolution of relative  $\ell_2$ errors and Pearson correlation coefficient of displacement magnitude $\left| \mathbf{x} \right|$ (right) for the test set.}
   \revdatanew{\textbf{e}, Snapshots of the reconstructed flow velocity magnitude fields (left) along with normalized absolute errors of predicted fields on the test set (right).} The central box spans the interquartile range from 25\% to 75\%, with the median marked by a horizontal line within the box. Lines extending from the box denote the 5\% to 95\% range. \revdatanew{ Sample sizes for the four time steps (from left to right) are $14,912$, $17,284$, $17,489$, and $17,043$ grid points.}     
}
   \label{fig:5}
\end{figure}

\subsection*{Experimental left ventricle blood flow}

\revdata{To further assess TrajectoryFlowNet's ability in particle trajectory tracking and flow field prediction with limited experimental data, we consider the flood flow dynamics within the left ventricle (LV) under physiological mitral valve conditions. }
This case presents a stringent test of the model's robustness due to the inherent noise and data incompleteness in real-world applications.

\rev{The blood flow patterns during LV filling are directly linked to diastolic efficiency, myocardial relaxation, and filling pressure. Abnormal filling patterns serve as key biomarkers of early heart failure~\cite{demirkiran2022left}. Combining noninvasive imaging with AI-based flow reconstruction enables the detection of subclinical diastolic dysfunction, supporting cardiac function assessment and early screening for heart dysfunction~\cite{smiseth2022multimodality,mele2025left}.}

The dataset was obtained by a lab experiment. The aortic root, aortic valve, and the complete mitral valve complex (including the annulus, anterior/posterior leaflets, chordae tendineae, and papillary muscles) were surgically dissected and sutured onto a 3D-printed annular resin plate connected to a silicone LV (see Fig. \ref{fig:5}\textbf{a}). The geometry of the silicone LV was reconstructed from adult heart CT data, with a wall thickness of 2 mm and a refractive index of 1.4. The mitral valve-silicone LV assembly was placed in a nonagonal chamber filled with a glycerol-water solution (62\% water, 38\% glycerol by volume), with a dynamic viscosity of 3.49 $\mathrm{mPa} \cdot \mathrm{s}$ at a temperature of 26.9$^\circ$C. A piston driven by an electric motor generated periodic compression and relaxation of the chamber fluid to simulate LV systole and diastole, at a heart rate of 60 beats per minute and a cardiac output of 3.6 L/min. The boundary deformation caused by the influence of the cardiac pulsation cycle is shown in Fig. \ref{fig:5}\textbf{b}. Similar to the previous case, time-resolved 2D PIV experiments were conducted to track the flow states. Particle image pairs were processed using an in-house PIV software, supplemented by manual tracking of particle trajectories. Further details about this dataset can be found in \textcolor{blue}{Supplementary Note C.4}. 


Fig. \ref{fig:5}\textbf{b}--\textbf{c} show the measured particle trajectories and sparse velocity field snapshots, respectively. The original training set contained 527 manually annotated particles with 4,711 spatiotemporal sample points, along with 14,133 single-time-step particles measured by PIV. Following the same data augmentation strategy as in the previous case, the long-range particle trajectories were partitioned into shorter trajectories with multiple temporal intervals (2-50 steps), where 18,000 sample points were randomly selected for training. 
\revdatanew{The test set was constructed in the same manner as discussed in the previous case, consisting of 96 manually annotated particles for trajectory evaluation (646 data points) and 200 PIV snapshots for velocity-field evaluation, all without overlap with the training data.}
The model was trained using L-BFGS.

\revdatanew{Fig. \ref{fig:5}\textbf{d} displays the predicted particle trajectories in test set and the corresponding relative  $\ell_2$ errors with the Pearson correlations of displacement magnitude $\left| \mathbf{x} \right|$ over time. Fig. \ref{fig:5}\textbf{e} shows the snapshots of the reconstructed flow velocity magnitudes, along with the normalized prediction errors computed exclusively on the test set at different time steps.} 
These results show that our model effectively captured the particle trajectories and the evolving flow structures within the LV.
Even under conditions of extremely sparse training data and unknown varying BCs, the Pearson correlation coefficients for predicted particle trajectories can still consistently achieve \revdatanew{0.98 as shown in Fig. \ref{fig:5}\textbf{d} (with the overall correlation values of predicted velocities above 0.73 as listed in Table \ref{table1})}. \rev{In addition, the evolution of the predicted dimensionless pressure fields with uncertainty quantification are shown in Extended Data Fig. \textcolor{blue}{3}\textbf{b}.}

We also observed that the prediction errors are positively correlated with the variations of the Reynolds number (see Extended Data Fig. \textcolor{blue}{4}). At the moment with a higher Reynolds number (e.g., at 0.5 s when the motor pumps the fluid resulting in a much larger flow velocity), the velocity errors increase significantly, as shown in Extended Data Fig. \textcolor{blue}{4}\textbf{a}. This is attributed to our use of a fixed Reynolds number (despite identifiable) in our model prediction. During periods when the Reynolds number varies rapidly, the model’s prediction performance degrades. Nevertheless, our results remain statistically meaningful, as shown in Fig. \ref{fig:5}\textbf{d}--\textbf{e}. Meanwhile, the velocity magnitude Pearson correlation remains stable over time (see Extended Data Fig. \textcolor{blue}{4}\textbf{b}), demonstrating its potential as a powerful tool for cardiovascular imaging and perhaps early-stage heart disease screening.

\section*{Discussion}
This paper introduces TrajectoryFlowNet, a model for fluid flow velocimetry and imaging under sparse data conditions. TrajectoryFlowNet predicts long-range particle trajectories from their initial positions and simultaneously estimates spatiotemporal velocity and pressure fields. The method is built on a Lagrangian–Eulerian physics-informed neural network that incorporates the continuity and Navier-Stokes equations as constraints, without the need of specifying the BCs. Consequently, the model effectively handles sparse, noisy data, generalizes to predict long-range trajectories of arbitrary particles, and accurately reconstructs the evolving flow field.

We evaluated the performance of TrajectoryFlowNet across multiple flow configurations: two numerical examples with synthetic data (lid-driven cavity flow, complex cylinder flow) and two experimental cases with real-world measurements (aortic blood flow and left ventricle blood flow).
\rev{Here, like most data-driven approaches, our model requires retraining for each different flow scenario.}
\rev{Nevertheless, within each trained case, TrajectoryFlowNet demonstrates strong robustness.}
Despite limited training data, our model effectively captured complex flow patterns and achieved high prediction accuracy. Notably, under uncertainties such as measurement noise, incomplete data, varying boundary conditions, and errors from model simplifications (e.g., assuming incompressible flow or neglecting blood aggregation), TrajectoryFlowNet demonstrated strong generalization performance. Overall, the method proves effective in learning spatiotemporal flow dynamics and accurately tracking long-term trajectories of passive particles.
\rev{Moreover, TrajectoryFlowNet holds promise for broader environmental and engineering applications, such as predicting ocean pollutant or microplastic dispersion from sparse drifter data, reconstructing Arctic surface currents using satellite-tracked sea ice floes, and recovering wake structures in wind tunnel tests from limited tracer particles.}

While TrajectoryFlowNet demonstrates strong generalization and accurate reconstruction, its current form faces several limitations that point to important future research directions. \rev{First, the model's point-wise neural architecture presents scalability challenges when dealing with large-scale, 3D, highly turbulent flows. Integrating graph neural networks to encode spatial connectivity and flow topology, along with turbulence-model-based Navier-Stokes equations as physical priors, might help resolve this issue, enabling more efficient and physically consistent learning in complex flow settings.} Second, although TrajectoryFlowNet exhibits resilience under measurement noise and variable conditions, it does not provide explicit uncertainty quantification for its predictions. A promising solution is to develop a probabilistic version of the framework (e.g., through variational Bayesian inference or diffusion models), which would enhance the model's reliability in real-world applications. Third, the present study is confined to 2D configurations. 
\rev{To enable 3D extension, the streamfunction formulation would be replaced by direct prediction of the velocity field, with incompressibility enforced via a divergence-free loss term.}
Extending the framework to three dimensions represents a great challenge. A key future goal is to leverage multi-view 2D particle-tracking data to establish a model capable of long-range 3D trajectory prediction along with 3D flow field reconstruction, a significant step toward practical application in complex flow environments.

\section*{Methods}\label{methods}

In this section, we introduce the methods of the proposed TrajectoryFlowNet model for simultaneous prediction of long-range particle trajectories and flow fields. Additional details are provided in \textcolor{blue}{Supplementary Note B}.

\subsection*{Lagrangian NS equations}
The proposed TrajectoryFlowNet aims to establish a unified learning model for flow velocimetry and imaging, realizing simultaneous prediction of long-range particle trajectories and spatiotemporal flow field. This framework creates a dynamic coupling mechanism between Lagrangian and Eulerian descriptions in data-driven simulation of complex flows. The network was designed based on the fundamental equivalence between fluid parcel trajectories and local flow states, where the velocity of a passive fluid parcel at any location directly corresponds to the Eulerian velocity at that point.

The Lagrangian trajectory of a $n$-dimensional fluid parcel (aka, a passive particle) can be parametrized by the spatiotemporal mapping $\mathbf{x}(\mathbf{x}_0,t_0,\tau) \in \mathbb{R}^n$. The parcel velocity, $\mathbf{u}\in \mathbb{R}^n$, equal to the Eulerian velocity at its current location, is given by:
\begin{equation}
    \mathbf{u}= \frac{d \mathbf{x}}{d \tau}(\mathbf{x}_0,t_0,\tau).
\end{equation}
In the Lagrangian description, the dynamics of a fluid parcel in the absence of body forces is governed by the momentum conservation law, expressed as:
\begin{equation}
    \rho \frac{D\mathbf{u}}{D\tau} + \boldsymbol{\nabla} p - \mu \boldsymbol{\nabla}^2 \mathbf{u} = \mathbf{0}.
\end{equation}
where $p\in \mathbb{R}$ denotes the pressure field, $\rho\in \mathbb{R}$ the flow density, and $\boldsymbol{\nabla} \in \mathbb{R}^n$ the Nabla operator.
Here,
\begin{equation}
    \frac{D\mathbf{u}}{D\tau} = \frac{\partial \mathbf{u}}{\partial t} + (\mathbf{u}\cdot\boldsymbol{\nabla})\mathbf{u},
\end{equation}
is the material derivative of velocity, which quantifies the time rate of velocity change following a fluid particle and naturally bridges the Lagrangian and Eulerian viewpoints.
Expanding the material derivative of velocity, the momentum equation is then obtained in the Eulerian form, given by:
\begin{equation}\label{NSeq}
\frac{\partial \mathbf{u}}{\partial t}+(\mathbf{u} \cdot \boldsymbol{\nabla}) \mathbf{u}=-\frac{1}{\rho} \boldsymbol{\nabla} p+\frac{\mu}{\rho} \boldsymbol{\nabla}^{2} \mathbf{u},
\end{equation}
along with the incompressibility condition (continuity equation) $\boldsymbol{\nabla} \cdot \mathbf{u} = 0$. Here, $\mu\in \mathbb{R}$ denotes the viscosity coefficient.

\subsection*{Network architecture}
To address the challenge of long-range particle trajectory tracking simultaneously along with full-field flow prediction, the architecture of TrajectoryFlowNet was designed to meet two key requirements. First, it must possess sufficient representation capacity to capture complex particle motions and flow dynamics across diverse flow regimes (e.g., transient behaviors, open or moving boundaries, and sparse, noisy observations). Second, its predictions should remain physical consistency with the governing equations of fluid motion. Guided by these two principles, we developed TrajectoryFlowNet (see Fig. \ref{fig:1}\textbf{b}), which simultaneously predicts long-range particle trajectories and reconstructs the corresponding spatiotemporal velocity and pressure fields. The framework consists of two major components: a Trajectory Block and a Flow Field Block. Together, these modules integrate the flexibility of data-driven learning with a hybrid Lagrangian-Eulerian representation in a unified physics-informed model, resulting in a robust performance.

\textbf{Trajectory Block.} In the Lagrangian frame, the Trajectory Block learns the spatiotemporal evolution of particle trajectories and forecasts their future motion given only their initial positions. The latent mapping is parameterized by a fully connected network $\mathcal{N}_1$, given by:
\begin{equation}
\tilde{\mathbf{x}} = \mathcal{N}_1(\mathbf{x}_0, t_0, \tau; \boldsymbol{\theta}),
\end{equation}
where $\boldsymbol{\theta}$ denotes the trainable parameters of $\mathcal{N}_1$. The network (see Fig. \ref{fig:1}\textbf{c}) propagates the input through multiple fully connected feedforward layers (with Tanh as the activation function) to predict the particle displacement $\tilde{\mathbf{x}}$. The output displacement is added element-wise to the initial position to update the particle trajectory $\mathbf{x}$, written as:
\begin{equation}
\mathbf{x} = \tilde{\mathbf{x}} + \mathbf{x}_0.
\end{equation}
This formulation is interpretable and enables the model with reliable long-term trajectory tracking in complex flow environments.

\textbf{Flow Field Block.} In the Eulerian frame, the Flow Field Block predicts the global velocity and pressure fields given the instantaneous particle locations. Unlike direct velocity regression, the network (see Fig. \ref{fig:1}\textbf{d}) outputs a stream function $\psi(\mathbf{x,t)}  \in \mathbb{R}$ and pressure $p$, ensuring that the predicted flow inherently satisfies the incompressibility condition.
\revnew{Although pressure recovery via the Poisson equation is theoretically feasible, it requires reliable boundary conditions that are often unavailable in cases with complex or moving boundaries. Meanwhile, enforcing the Poisson equation through automatic differentiation introduces greater sensitivity to errors compared to direct pressure prediction. Therefore, we adopt the direct pressure prediction approach in our framework.}
Specifically, the latent variables $\{\psi, p\}$ are generated by a fully connected network $\mathcal{N}_2$ parameterized by $\boldsymbol{\vartheta}$:
\begin{equation}
\{\psi, p\} = \mathcal{N}_2(\mathbf{x}, t; \boldsymbol{\vartheta}).
\end{equation}

The velocity components (e.g., for a 2D flow field) are then derived from the spatial derivatives of the stream function by automatic differentiation:
\begin{equation}
u = \frac{\partial\psi}{\partial y}, \quad v= - \frac{\partial\psi}{\partial x},
\end{equation}
which eliminates the need for explicit loss term for $\boldsymbol{\nabla} \cdot \mathbf{u} = 0$ during training. Similar to $\mathcal{N}_1$, the inputs are processed through multiple feedforward layers with Tanh activations to capture smooth, continuous variations in velocity and pressure across space and time. 
\rev{Additionally, boundary conditions are not explicitly imposed when training TrajectoryFlowNet, but are implicitly constrained by the spatial envelope of observed particle trajectories and velocities.}

\textbf{Normalization.} To ensure stable optimization and consistent physical scaling, we employ a normalization strategy inspired by classical nondimensionalization in fluid mechanics. All physical quantities are transformed into dimensionless forms using characteristic reference scales. Specifically, spatial coordinates and velocities are normalized by their respective maximal reference values $L$ and $U$, expressed as:
\begin{equation}
 \mathbf{x} \leftarrow \frac{\mathbf{x}}{L}, \quad
 \mathbf{x}_0 \leftarrow \frac{\mathbf{x}_0}{L}, \quad
 \mathbf{u} \leftarrow \frac{\mathbf{u}}{U}, \quad 
 p \leftarrow \frac{p}{\rho U^2}, \quad 
 t \leftarrow \frac{t U}{L} , \quad
 t_0 \leftarrow \frac{t_0 U}{L} , \quad
 \tau \leftarrow \frac{\tau U}{L}. 
\end{equation}
Under this normalization, the incompressible Navier–Stokes equations become:
\begin{equation}
 \frac{\partial \mathbf{u}}{\partial t} + (\mathbf{u} \cdot {\boldsymbol{\nabla}}) \mathbf{u} + {\boldsymbol{\nabla}} p - \frac{1}{\text{Re}} {\boldsymbol{\nabla}}{^2} \mathbf{u}=\mathbf{0}, 
\end{equation}
where the variables in the above equation are dimensionless and $\text{Re} = \rho U L/\mu$ denotes the Reynolds number, which is trainable if is unknown. This nondimensionalization confines all input and output quantities within the interval $[-1, 1]$, aligning naturally with the range of the Tanh activation function. Such an approach promotes numerical stability, accelerates network convergence, and ensures physically consistent scaling across diverse flow conditions.

\textbf{Fourier feature mapping.} A major challenge in tracking large-deformation particle trajectories and modeling complex flow phenomena (e.g., turbulent flows) is how to enhance the network’s representation capacity without excessively increasing trainable parameters. To this end, we incorporate Fourier feature mapping into both network modules. 
\rev{This technique projects the input coordinates onto a higher-dimensional hypersphere using sinusoidal basis functions, enabling the neural network to approximate high-frequency variations within low-dimensional inputs. Prior studies have shown that such encoding not only facilitates the representation of periodic structures~\cite{dong2021method} but also alleviates spectral bias in multi-scale problems by reshaping the Neural Tangent Kernel spectrum~\cite{wang2021eigenvector}.}
The mapping is given by:
\begin{equation}
\boldsymbol{\gamma}(\mathbf{v})=[\cos(2\pi\mathbf{Bv}), \sin(2\pi\mathbf{Bv)}]^\mathrm{T},
\end{equation}
where, $\mathbf{B}$ is a Gaussian random matrix with elements $B_{ij} \sim\mathcal{N}(0,\sigma_{ffm}^2)$, $\sigma_{ffm}$ a hyperparameter tuned according to different tasks, and $\mathbf{v}$ the input vectors to be embedded (e.g. ($\mathbf{x}_0, t_0, \tau$) or ($\mathbf{x},t$)).

For the Trajectory Block, Fourier feature mapping is applied to the Lagrangian inputs $(\mathbf{x}_0, t_0, \tau)$, expressed as:
\begin{equation}
\boldsymbol{\gamma}_1=\boldsymbol{\gamma}_1(\mathbf{x_0}, t_0, \tau; \mathbf{B}_1),
\end{equation}
where $\mathbf{B}_1$ denotes the random projection matrix specific to this block.
Similarly, for the Flow Field Block, the Eulerian spatiotemporal coordinates $(\mathbf{x},t)$ are transformed through Fourier features generated by an independent Gaussian random matrix $\mathbf{B}_2$, given by:
\begin{equation}
\boldsymbol{\gamma}_2 = \boldsymbol{\gamma}_2(\mathbf{x}, t; \mathbf{B}_2).
\end{equation}
These mappings enrich the input representations, allowing TrajectoryFlowNet to capture fine-scale variations in both particle trajectories and fluid fields while maintaining model compactness. Specific parameter choices and implementation details are provided in \textcolor{blue}{Supplementary Note D.1}.

\textbf{Loss functions.} Accurate flow reconstruction requires that the network’s predictions remain consistent with underlying physical laws. Accordingly, a physics-constrained loss function is introduced, combining data-driven supervision with physical regularization, as shown in Fig. \ref{fig:1}. The total loss $\mathcal{L}$ consists of two components: the data loss $\mathcal{L}_{d}$ and the physics loss $\mathcal{L}_{p}$.

The data loss $\mathcal{L}_{d}$ evaluates the discrepancy between predicted and reference trajectories and velocity fields, consisting of two terms, namely, $\mathcal{L}_x(\boldsymbol{\theta};\mathcal{D}_d)$ for trajectories and $\mathcal{L}_v(\boldsymbol{\theta},\boldsymbol{\vartheta}; \mathcal{D}_d)$ for the velocity field, defined as:
\begin{equation}
 \mathcal{L}_x(\boldsymbol{\theta}; \mathcal{D}_d) =\frac{1}{N_d}\sum_{i=1}^{N_d}  \|\hat{\mathbf{x}}^{i} - {\mathbf{x}}^i \|_2^2,
\end{equation}
\begin{equation}
 \mathcal{L}_v(\boldsymbol{\theta}, \boldsymbol{\vartheta}; \mathcal{D}_d) =  \frac{1}{N_d} \sum_{i=1}^{N_d} \|\hat{\mathbf{u}}^{i} - {\mathbf{u}}^i \|_2^2,  
\end{equation}
where $\hat{\mathbf{x}}^i\in \mathbb{R}^{n} $ and $\hat{\mathbf{u}}^i\in \mathbb{R}^{n} $ denote the predicted particle trajectories and velocity fields of the $i$th spatiotemporal points ($i=1, 2, ..., N_d$), $\{\mathbf{x}^i, \mathbf{u}^i\}$ the corresponding ground-truth data, $N_d$ the number of training data points, and $\|\cdot\|_2 $ the Frobenius norm. These terms drive the network parameters ($\boldsymbol{\theta}, \boldsymbol{\vartheta}$) toward data fidelity across both Lagrangian and Eulerian representations.

To impose physical consistency, the physics loss $\mathcal{L}_p$ minimizes violations of the governing equations, consisting of the trajectory-velocity compatibility loss $\mathcal{L}_{com}(\boldsymbol{\theta},\boldsymbol{\vartheta}; \mathcal{D}_c)$ and the NS equation loss $\mathcal{L}_{NS}(\boldsymbol{\theta},\boldsymbol{\vartheta}; \mathcal{D}_c)$, written as:
\begin{equation}
\mathcal{L}_{com}(\boldsymbol{\theta}, \boldsymbol{\vartheta}; \mathcal{D}_c) = \frac{1}{N_c}\sum_{j=1}^{N_c} \left\| \left(\frac{d\hat{\mathbf{x}}^{j}}{d\tau^{j}} \right) - \hat{\mathbf{u}}^{j} \right\|_2^2,
\end{equation}
\begin{equation}
\mathcal{L}_{NS}(\boldsymbol{\theta}, \boldsymbol{\vartheta}; \mathcal{D}_c) =  \frac{1}{N_c} \sum_{j=1}^{N_c}  \left\| \left(\frac{d\hat{\mathbf{u}}^{j}}{dt^{j}} + (\hat{\mathbf{u}}^{j} \cdot {\boldsymbol{\nabla}}) \hat{\mathbf{u}}^{j} + {\boldsymbol{\nabla}} \hat{p}^{j} - \frac{1}{\hat{\text{Re}}} {{\boldsymbol{\nabla}}}^2 \hat{\mathbf{u}}^{j} \right) \right\|_2^2, 
\end{equation}
where $\hat{\mathbf{x}}^j\in \mathbb{R}^{n} $, $\hat{\mathbf{u}}^j\in \mathbb{R}^{n} $ and   $\hat{p}^j\in \mathbb{R} $ represent the predicted trajectories, velocity and pressure of the $j$th spatiotemporal points ($ij=1, 2, ..., N_c$), respectively; $\hat{\text{Re}}\in \mathbb{R} $ denote the Reynolds number (treated as an additional trainable parameter if unknown), $N_c$ the number of collocation data points used to enforce the physical constraints. 
\revdatanew{
In summary, the total loss function is written as:
\begin{equation}
    \mathcal{L}(\boldsymbol{\theta}, \boldsymbol{\vartheta}; \mathcal{D}_d, \mathcal{D}_c) = \lambda_1\mathcal{L}_x(\boldsymbol{\theta};\mathcal{D}_d) + \lambda_2\mathcal{L}_v(\boldsymbol{\theta}, \boldsymbol{\vartheta}; \mathcal{D}_d) + \lambda_3\mathcal{L}_e(\boldsymbol{\theta}, \boldsymbol{\vartheta}; \mathcal{D}_c) + 
    \lambda_4\mathcal{L}_{NS}(\boldsymbol{\theta}, \boldsymbol{\vartheta}; \mathcal{D}_c),
\end{equation}
where $\lambda_1$, $\lambda_2$, $\lambda_3, \lambda_4$ are the loss hyperparameters. }
\rev{Following common practice in PINNs, the loss hyperparameters $\lambda_i$ are empirically tuned. We note that model performance can be sensitive to the choice of $\lambda_i$, and all selected values are provided in the open-source code repository for reproducibility.}
By jointly minimizing $\mathcal{L}_d$ and $\mathcal{L}_p$, the model learns flow representations that are simultaneously data-physics consistent, alleviating overfitting issues while improving the model's generalizability to unseen flow conditions.

\subsection*{Evaluation metrics}
We define five metrics to quantitatively evaluate the prediction accuracy of our model, including the Pearson correlation, relative  $\ell_2$ error, root mean square error (RMSE), mean absolute error (MAE) and normalized absolute error (NAE), described as follows.

The Pearson correlation quantifies the linear alignment between the predicted solution $\mathbf{u}_{pre}$ and the reference solution $\mathbf{u}$, defined as:
\begin{equation}
\text{Correlation}(\mathbf{u}_{pre},\mathbf{u}) = \frac{\sum_{i=1}^{N} (\mathbf{u}_{pre}^i - \bar{\mathbf{u}}_{pre})(\mathbf{u}^i - \bar{\mathbf{u}})}{\sqrt{\sum_{i=1}^{N} (\mathbf{u}_{{pre}}^i - \bar{\mathbf{u}}_{pre})^2}  \sqrt{\sum_{i=1}^{N} (\mathbf{u}^i - \bar{\mathbf{u}})^2}},
\end{equation}
where $i$ denotes the sample index, $N$ the number of the total data points, $\bar{\mathbf{u}}_{pre}$ and $\bar{\mathbf{u}}$ the mean values of $\mathbf{u}_{pre}$ and $\mathbf{u}$, respectively.

The relative  $\ell_2$ error measures the normalized difference between the predicted solution $\mathbf{u}_{pre}$ and the reference solution $\mathbf{u}$, given by:
\begin{equation}
\text{Relative  $\ell_2$ error} = \frac{\| \mathbf{u} - {\mathbf{u}_{pre}} \|_2}{\| \mathbf{u} \|_2}.
\end{equation}
where $\|\cdot\|$ denotes the Euclidean norm.

The RMSE quantifies the average magnitude of the prediction errors, defined as:
\begin{equation}
\text{RMSE} = \sqrt{\frac{1}{N} \sum_{i=1}^{N} \left(\mathbf{u}^i - \mathbf{u}_{pre}^i\right)^2}.
\end{equation}

The MAE measures the average absolute difference between the predicted and ground-truth values, given by:
\begin{equation}
\text{MAE} = \frac{1}{N} \sum_{i=1}^{N} |\mathbf{u}^i - \mathbf{u}_{pre}^i|.
\end{equation}

The normalized absolute error enables scale-invariant comparison by normalizing the absolute deviations using the maximum absolute value of the ground truth as the reference scale, namely,
\begin{equation}
\epsilon = \frac{|\mathbf{u}^i - \mathbf{u}_{pre}^i|}{ |\tilde{\mathbf{u}}|},
\end{equation}
where $\tilde{\mathbf{u}}$ denotes the characteristic velocity (e.g., 0.5 m/s in experimental cases).

\subsection*{Experimental setup and computational resources}
\revdatanew{The models were trained using the L-BFGS optimizer with a learning rate of 1.0 and a maximum of 500,000 iterations (unless otherwise specified). For the lid‑driven cavity flow case, the Trajectory Block adopted a 4‑layer MLP with 40 neurons per hidden layer, and the Flow Field Block used a 6‑layer MLP with 60 neurons per hidden layer. For the complex cylinder flow case, the Trajectory Block consisted of 10 hidden layers with 40 neurons each, whereas the Flow Field Block remained a 6‑layer MLP with 60 neurons per hidden layer, with other configurations kept identical to the lid‑driven cavity case. For the experimental cases, the Trajectory Block employed a 4‑layer MLP (40 neurons per hidden layer), and the Flow Field Block was set to 10 hidden layers with 60 neurons per layer.} Model training was conducted on a server equipped with an Intel(R) Xeon(R) Platinum 8380 CPU (2.30GHz, 64 cores) and a NVIDIA A100 GPU (80GB).

\section*{Data availability} 
All the used datasets \yledit{(the synthetic lid-driven cavity flow data, the synthetic complex cylinder flow data, the experimental aortic blood flow data, and the experimental left ventricle blood flow data)} in this study are available on GitHub at \url{https://github.com/WanJingdi/TrajectoryFlowNet} \cite{wan_2026_20252180}.

\section*{Code availability} 
All the source codes to reproduce the results in this study are available on GitHub at \url{https://github.com/WanJingdi/TrajectoryFlowNet} \cite{wan_2026_20252180}.

\vspace{36pt}
\section*{Funding Statement}
The work is supported by the National Natural Science Foundation of China (No. 62576331 and No. 12588201), the National Key R\&D Program of China (No. 2025ZD0122000), the Strategic Priority Research Program of the Chinese Academy of Sciences (No. XDB0620103), and the Fundamental Research Funds for the Central Universities (No. E2EG2202X2). The funders had no role in study design, data collection and analysis, decision to publish or preparation of the manuscript.

\section*{Author Contributions Statement} 
J.W. and Y.L. contributed to the ideation and design of the research; J.W., H.W., B.L., X.Y., X.H., S.C., and Y.L. performed the research; G.H. and Y.L. supervised the work; all authors contributed to research discussions, writing and editing the paper. 

\section*{Corresponding Author} 
Yang Liu (\url{liuyang22@ucas.ac.cn}).

\section*{Competing Interests Statement}
The authors declare no competing interests.

\section*{Supplementary information:}
The supplementary information is attached.



\clearpage
\setcounter{figure}{0}
\renewcommand{\figurename}{Extended Data Figure}
\setcounter{table}{0}
\renewcommand{\tablename}{Extended Data Table}


\begin{figure}[htbp]
  \centering
   \includegraphics[width=1.0\linewidth]{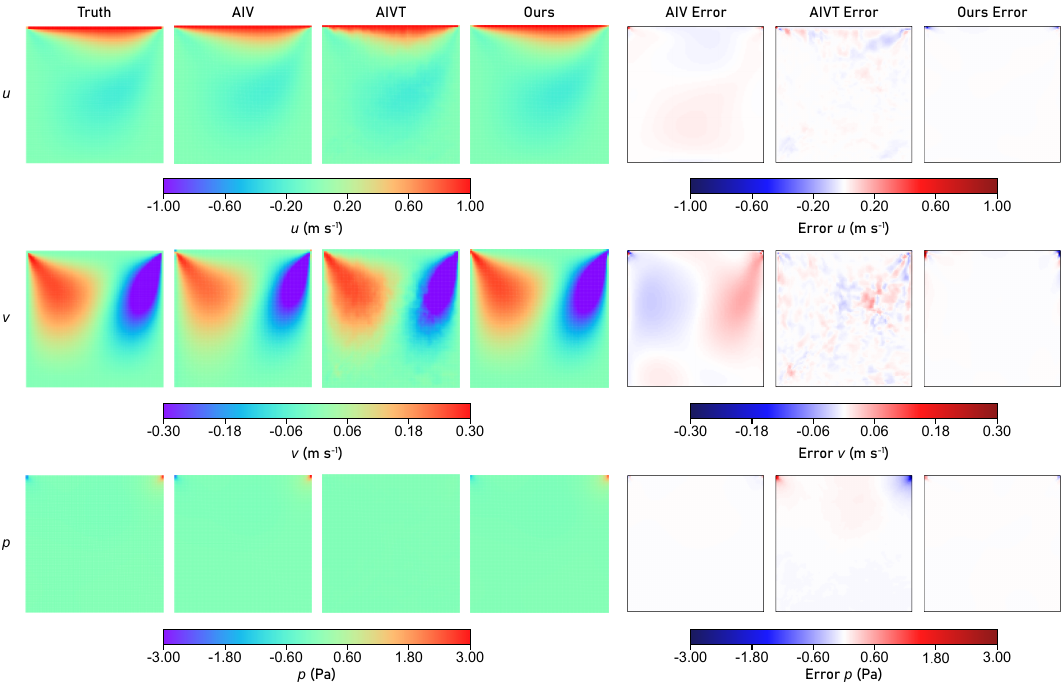}
   \caption{
   \textbf{Comparison of flow field prediction $(u,v,p)$ at $t = 25$ s in the lid-driven cavity flow using TrajectoryFlowNet, AIV and AIVT.}
   }
   \label{fig:Compare2AIVCase1}
\end{figure}

\clearpage

\begin{figure}[htbp]
  \centering
   \includegraphics[width=1.0\linewidth]{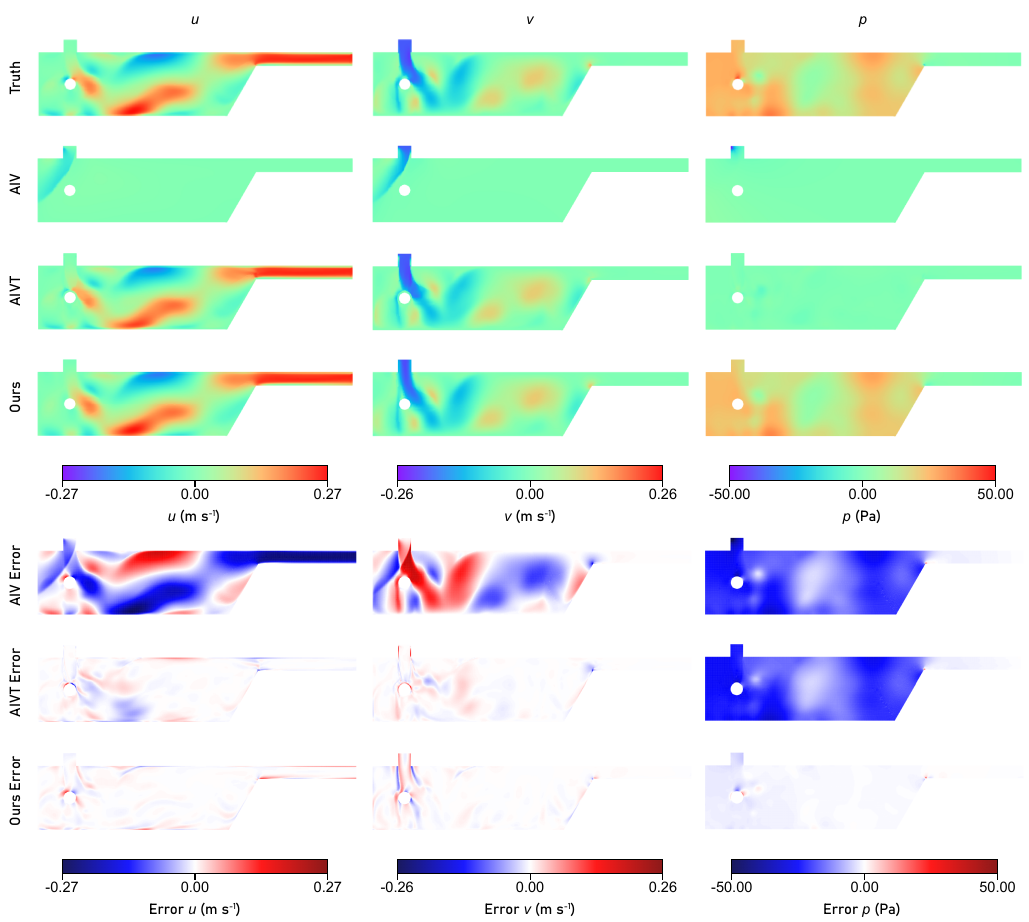}
   \caption{
   \textbf{Comparison of flow field prediction $(u,v,p)$ at $t = 25$ s in the complex cylinder flow using TrajectoryFlowNet, AIV and AIVT.}
   }
   \label{fig:Compare2AIVCase2}
\end{figure}

\begin{table}[htbp]
\setlength{\tabcolsep}{20pt}
\centering
\caption{\rev{Model performance comparison in terms of Pearson correlation (two-sided) at $t = 25$ s.}}
\label{AIVAIVT}
\small
\begin{tabular}{@{}lccccc@{}}
\toprule
\multirow{2}{*}{Verifiable Example} & \multirow{2}{*}{Model} & \multicolumn{3}{c}{Pearson Correlation} \\
\cmidrule(l){3-5}
& & $u$ & $v$ & $p$ \\
\midrule
\multirow{3}{*}{Lid-driven cavity flow} 
& AIV  & 0.9982 & 0.9987  & \pmb{0.9941}  \\
& AIVT   & 0.9973 & 0.9983  & 0.4934  \\
& Ours  & \pmb{0.9991} & \pmb{0.9989}  & 0.9898 \\
\midrule
\multirow{3}{*}{Complex cylinder flow}
& AIV  & 0.0586 & 0.3335  & 0.1228  \\
& AIVT   & 0.9924 & \pmb{0.9922}  & 0.3255 \\
& Ours  & \pmb{0.9959} & 0.9902  & \pmb{0.9943}\\

\bottomrule

\end{tabular}
\end{table}

\clearpage
\begin{figure}[htbp]
  \centering
   \includegraphics[width=0.8\linewidth]{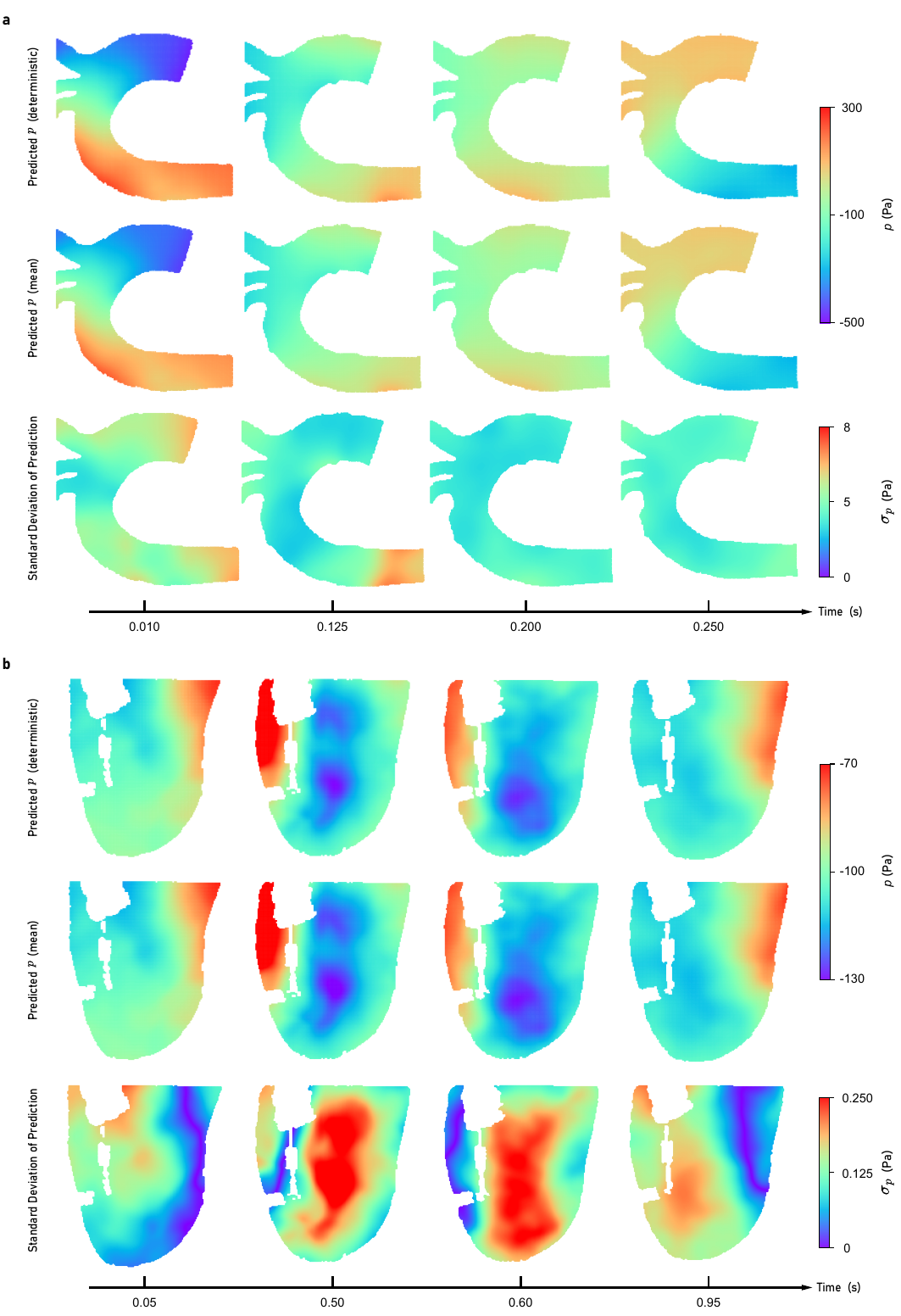}
   \caption{
   \textbf{Predicted pressure fields (p) of TrajectoryFlowNet in experimental cases.}
   \textbf{a}, Pressure predictions for experimental aortic blood flow.
   \textbf{b}, Pressure predictions for experimental left ventricle blood flow. 
   Here, “predicted p (deterministic)” denotes the prediction by deterministic TrajectoryFlowNet, “predicted p (mean)” the mean pressure field predicted by probabilistic TrajectoryFlowNet with 100 Stein particles, and the “standard deviation of prediction” quantifies the prediction uncertainties.
   }
   \label{fig:ExpPressure}
\end{figure}

\clearpage
\begin{figure}[htbp]
  \centering
   \includegraphics[width=1.0\linewidth]{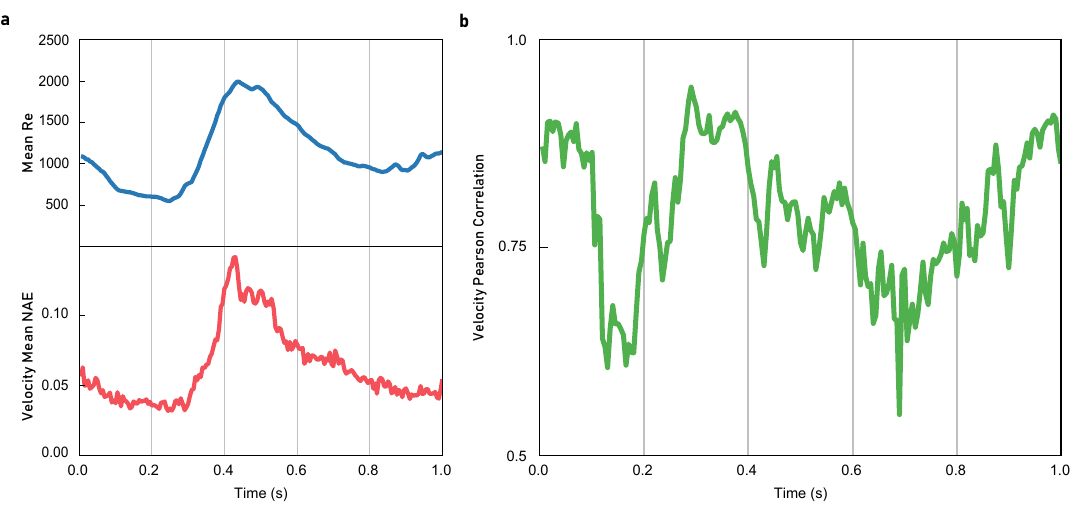}
   \caption{
   \textbf{Time-varying results of experimental left ventricle blood flow.}
   \textbf{a}, Variation of mean Reynolds number and velocity magnitude mean NAE over time, with $\rho =1328.4kg/m^{3}$, $L=23.0mm$ , and $\mu =3.345\times {10}^{-3}Pa\cdot s$.
   \textbf{b}, Variation of velocity Pearson correlation over time.
   }
   \label{fig:ReandError}
\end{figure}

\end{document}









{\footnotesize
\tableofcontents
}
\newpage

\noindent This supplementary document provides a detailed description of the proposed TrajectoryFlowNet model, examples, and additional test results.

\section{Background}
The research on tracking and quantifying flow fields aims to investigate and elucidate the formation, instability, and predictive capabilities of complex flow states. Predicting long-range motion paths of passive particles within a flow, while simultaneously inferring the flow's velocity and pressure is crucial for nature and engineering applications. The partial differential nature of the Navier-Stokes (NS) equations makes solving these problems particularly challenging. In recent years, machine learning has shown great promise in solving fluid mechanics issues described by partial differential equations. In this paper, we propose TrajectoryFlowNet, a Lagrangian–Eulerian physics-informed neural network architecture, for fluid flow velocimetry and imaging via learning to predict the spatiotemporal flow field and long-range trajectory of any arbitrary passive particle. Basic information on the fully connected neural network used in this study is presented below.

\textbf{Fully Connected Neural Network.} A fully connected neural network (FCNN)~\cite{lecun2015deep} is a type of artificial neural network that is used for solving various machine learning tasks. It consists of multiple layers of interconnected nodes or neurons, where each neuron in one layer is connected to all the neurons in the next layer.
The architecture of an FCNN can be represented mathematically by a set of weight matrices $W_l$, where $l$ denotes the layer index. Each weight matrix represents the connections between the neurons in the previous layer and the current layer. The input to the first layer is usually a vector $x\in R^n$ representing the features of the data sample. The output of the $i$th neuron in the $l$th layer can be calculated as:
\begin{equation}
z^{l}_i = \sigma_{act}\left(\sum_{j=1}^n w^{l}_{ij}a^{l-1}_j + b^{l}_i\right),
\label{eq:1} 
\end{equation}
where $\sigma_{act}$ is the activation function, $a^{l-1}_j$ is the output of the $j$th neuron in the previous layer, $w^{l}_{ij}$ is the weight connecting the $j$th neuron in the previous layer to the $i$th neuron in the current layer, and $b^{l}_i$ is the bias term associated with the $i$th neuron in the current layer.
The output of the last layer is usually interpreted as the prediction of the model for the given input. 

\section{Methodology}
\subsection{Algorithm description}
We provide a pseudo-code describing our proposed model, as shown in the following Algorithm~\ref{Algorithm 1}.
\begin{breakablealgorithm}
    \caption{TrajectoryFlowNet}
    \label{Algorithm 1} 
    \begin{algorithmic}[htbp]
    \smaller
        \State Initialize parameters $\boldsymbol{\theta}$ of the fully connected network $\mathcal{N}_1$
        \State Initialize parameters $\boldsymbol{\vartheta}$ of the fully connected network $\mathcal{N}_2$
        \State Define the training dataset: $\{(x_0^i, y_0^i, t_0^i, \tau^i, x^i, y^i, u^i, v^i)\}_{i=1}^{N_d}$
        \State Define collocation points: $\{(x_0^j, y_0^j, t_0^j, \tau^j)\}_{j=1}^{N_c}$
        \State Define Gaussian random matrix: $\{ \mathbf{B}_1, \mathbf{B}_2\}$
        \While{\text{not converged}}
            \For{each data point $({x_0}^i, {y_0}^i, {t_0}^i, {\tau}^i)$}
                \State Compute Fourier features:
                \[
                \boldsymbol{\gamma}_1(\mathbf{x}_0^d) = [\cos(2\pi \mathbf{B}_1 \mathbf{x}_0^d), \sin(2\pi \mathbf{B}_1 \mathbf{x}_0^d)]^\top
                \]
                where $\mathbf{x}_0^d = [{x_0}^i, {y_0}^i, {t_0}^i, {\tau}^i]^\top$
                \State Predict particle displacement $({\tilde{x}^{i}}, {\tilde{y}^{i}})$ using network $\mathcal{N}_1$:
                \[
                ({\tilde{x}^{i}}, {\tilde{y}^{i}}) = \mathcal{N}_1(\boldsymbol{\gamma}_1(\mathbf{x}_0^d); \boldsymbol{\theta})
                \]
                \State Compute particle new positions:
                \[
                \hat{x}^{i} = {x_0}^i + {\tilde{x}^{i}}, \quad\hat{y}^{i} = {y_0}^i + {\tilde{y}^{i}}
                \]
                \State Compute motion time under Euler description:
                \[
                t^{i} = {t_0}^i + {\tau^{i}}
                \]
                \State Compute Fourier features:
                \[
                \boldsymbol{\gamma}_2(\mathbf{x}^d) = [\cos(2\pi \mathbf{B}_2 \mathbf{x}^d), \sin(2\pi \mathbf{B}_2 \mathbf{x}^d)]^\top
                \]
                where $\mathbf{x}^d = [\hat{x}^{i}, \hat{y}^{i}, {t}^i]^\top$
                \State Predict $(\psi^{i}, \hat{p}^{i})$ using network $\mathcal{N}_2$:
                \[
                (\psi^{i}, \hat{p}^{i}) = \mathcal{N}_2(\boldsymbol{\gamma}_2(\mathbf{x}^d); \boldsymbol{\vartheta})
                \]
                \State Compute velocity $(\hat{u}^{i}, \hat{v}^{i})$:
                \[
                \hat{u}^{i} = \frac{\partial \psi^{i}}{\partial {{\hat{y}^{i}}}}, \quad 
                \hat{v}^{i} = -\frac{\partial \psi^{i}}{\partial {{\hat{x}^{i}}}}
                \]                
            \EndFor
            
            \For{each collocation point $({x_0}^j, {y_0}^j, {t_0}^j, {\tau}^j)$}
                \State Compute Fourier features:
                \[
                \boldsymbol{\gamma}_1(\mathbf{x}_0^c) = [\cos(2\pi \mathbf{B}_1 \mathbf{x}_0^c), \sin(2\pi \mathbf{B}_1 \mathbf{x}_0^c)]^\top
                \]
                where $\mathbf{x}_0^c = [{x_0}^j, {y_0}^j, {t_0}^j, {\tau}^j]^\top$
                
                \State Predict particle displacement $(\tilde{x}^{j}, \tilde{y}^{j})$ using network $\mathcal{N}_1$:
                \[
                (\tilde{x}^{j}, \tilde{y}^{j}) = \mathcal{N}_1(\boldsymbol{\gamma}_1(\mathbf{x}_0^c); \boldsymbol{\theta})
                \]
                \State Compute particle new positions:
                \[
                \hat{x}^{j} = {x_0}^j + {\tilde{x}^{j}}, \quad \hat{y}^{j} = {y_0}^j + {\tilde{y}^{j}}
                \]
                \State Compute motion time under Euler description:
                \[
                t^{j} = {t_0}^j + {\tau^{j}}
                \]
                \State Compute Fourier features:
                \[
                \boldsymbol{\gamma}_2(\mathbf{x}^c) = [\cos(2\pi \mathbf{B}_2 \mathbf{x}^c), \sin(2\pi \mathbf{B}_2 \mathbf{x}^c)]^\top
                \]
                where $\mathbf{x}^c = [\hat{x}^{j}, \hat{y}^{j}, {t}^j]^\top$
                \State Predict $(\psi^{j}, \hat{p}^{j})$ using network $\mathcal{N}_2$:
                \[
                (\psi^{j}, \hat{p}^{j}) = \mathcal{N}_2(\boldsymbol{\gamma}_2(\mathbf{x}^c); \boldsymbol{\vartheta})
                \]
                \State Compute velocity $(\hat{u}^{j}, \hat{v}^{j})$:
                \[
                \hat{u}^{j} = \frac{\partial \psi^{j}}{\partial {{\hat{y}^{j}}}}, \quad 
                \hat{v}^{j} = -\frac{\partial \psi^{j}}{\partial {{\hat{x}^{j}}}}
                \]    
                \State Compute the time derivative of $\hat{x}^{j}$ and $\hat{y}^{j}$ under Lagrangian description:
                \[
                \frac{\partial \hat{x}^{j}}{\partial {\tau}^j}, \quad \frac{\partial \hat{y}^{j}}{\partial {\tau}^j}
                \]
                \State Compute the time derivative of $\hat{u}^{j}$ and $\hat{v}^{j}$  under Euler description:
                \[
                \frac{\partial \hat{u}^{j}}{\partial {t}^j}, \quad \frac{\partial \hat{v}^{j}}{\partial {t}^j}
                \]
                \State Compute the position derivative of $\hat{u}^{j}$ and $\hat{v}^{j}$:
                \[
                \frac{\partial \hat{u}^{j}}{\partial {{\hat{x}^{j}}}}, \quad 
                \frac{\partial \hat{u}^{j}}{\partial {{\hat{y}^{j}}}}, \quad 
                \frac{\partial^2 \hat{u}^{j}}{\partial {{\hat{x}^{j2}}}}, \quad 
                \frac{\partial^2 \hat{u}^{j}}{\partial {{\hat{y}^{j2}}}}, \quad 
                \frac{\partial \hat{v}^{j}}{\partial {{\hat{x}^{j}}}}, \quad 
                \frac{\partial \hat{v}^{j}}{\partial {{\hat{y}^{j}}}}, \quad 
                \frac{\partial^2 \hat{v}^{j}}{\partial {{\hat{x}^{j2}}}}, \quad 
                \frac{\partial^2 \hat{v}^{j}}{\partial {{\hat{y}^{j2}}}}
                \]
                \State Compute the position derivative of $\hat{p}^{j}$:
                \[
                \frac{\partial \hat{p}^{j}}{\partial {{\hat{x}^{j}}}}, \quad 
                \frac{\partial \hat{p}^{j}}{\partial {{\hat{y}^{j}}}}
                \]
                \State Compute the residuals of the non-dimensional Navier-Stokes equations:
                \[
                R_x = \frac{\partial \hat{u}^{j}}{\partial t^{j}} + \hat{u}^{j} \frac{\partial \hat{u}^{j}}{\partial {{\hat{x}^{j}}}} + \hat{v}^{j} \frac{\partial \hat{u}^{j}}{\partial {{\hat{y}^{j}}}} - \frac{1}{\text{Re}} \left( \frac{\partial^2 \hat{u}^{j}}{\partial {{\hat{x}^{j2}}}} + \frac{\partial^2 \hat{u}^{j}}{\partial {{\hat{y}^{j2}}}} \right) + \frac{\partial \hat{p}^{j}}{\partial {{\hat{x}^{j}}}}
                \]
                \[
                R_y = \frac{\partial \hat{v}^{j}}{\partial t^{j}} + \hat{u}^{j} \frac{\partial \hat{v}^{j}}{\partial {{\hat{x}^{j}}}} + \hat{v}^{j} \frac{\partial \hat{v}^{j}}{\partial {{\hat{y}^{j}}}} - \frac{1}{\text{Re}} \left( \frac{\partial^2 \hat{v}^{j}}{\partial {{\hat{x}^{j2}}}} + \frac{\partial^2 \hat{v}^{j}}{\partial {{\hat{y}^{j2}}}} \right) + \frac{\partial \hat{p}^{j}}{\partial {{\hat{y}^{j}}}}
                \]
            \EndFor
            \State Compute total loss:
            \[
            \mathcal{L} =  \frac{1}{N_d} \sum_{i=1}^{N_d} \left( \left| \hat{x}^{i}  - x^{i} \right|^2 + \left| \hat{y}^{i}  - y^{i} \right|^2 + \left| \hat{u}^{i} - u^{i} \right|^2 + \left| \hat{v}^{i} - v^{i} \right|^2 \right)
            \]  
            \[
            + \frac{1}{N_c} \sum_{j=1}^{N_c} \left( \left| \frac{\partial \hat{x}^{j}}{\partial {\tau}^j} - \hat{u}^{j} \right|^2 + \left| \ \frac{\partial \hat{y}^{j}}{\partial {\tau}^j} - \hat{v}^{j} \right|^2 + |R_x|^2 + |R_y|^2 \right)
            \]
            \State Update the parameters of both networks with learning rate $\eta$:
            \[
            \boldsymbol{\theta} \leftarrow \boldsymbol{\theta} - \eta \boldsymbol{\nabla}_{\boldsymbol{\theta}} \mathcal{L}, \quad \boldsymbol{\vartheta} \leftarrow \boldsymbol{\vartheta} - \eta \boldsymbol{\nabla}_{\boldsymbol{\vartheta}} \mathcal{L}
            \]
        \EndWhile
        \Return the predicted values  $(\hat{x}, \hat{y}, \hat{u}, \hat{v}, \hat{p})$
    \end{algorithmic}
\end{breakablealgorithm} 

\subsection{Variables notation}
We present a summary of the variable notations used in our paper, as detailed in Table \ref{table1}.

\begin{table}[h!] 
\centering 
\caption{Variables notation used in our paper.} 
\label{table1} 
\begin{tabular}{@{}lccc@{}} 
\toprule 
Variable Name & Short Name & Role \\
\midrule 
Space coordinate's initial x-position of the particle & $x_0$ & Input \\
Space coordinate's initial y-position of the particle & $y_0$ & Input \\
Particle's initial motion time under Euler description & $t_0$ & Input \\
Particle's motion time under Lagrangian description & $\tau$ & Input \\
Particle's motion time under Euler description & $t$ & Input \\
Reynolds number & $\text{Re}$ & Input/Predicted \\
Particle's x-direction displacement of space coordinate & $\tilde{x}$ & Predicted \\ 
Particle's y-direction displacement of space coordinate & $\tilde{y}$ & Predicted \\ 
Particle's x-direction of space coordinate & $x$ & Input/Predicted \\ 
Particle's y-direction of space coordinate & $y$ & Input/Predicted \\ 
Velocity of x-component & $u$ & Input/Predicted \\ 
Velocity of y-component & $v$ & Input/Predicted \\ 
Stream function & $\psi$ & Predicted \\
Pressure & $p$ & Predicted \\
\bottomrule 
\end{tabular}
\end{table}

\section{Dataset}
We evaluated the performance of our proposed method using two simulated datasets and two real experimental dataset. The simulated datasets were generated using ANSYS FLUENT, featuring a lid-driven cavity flow with closed boundaries and a complex cylinder flow characterized by open boundaries. The experimental dataset comprised measurements of aortic blood flow and left ventricle blood flow were obtained from laboratory experiments. Each dataset was divided into separate training and testing sets to build the model. The model was trained on the training dataset and evaluated on the unseen testing dataset to assess its performance. 


\begin{figure}[t!]
  \centering
   \includegraphics[width=1.0\linewidth]{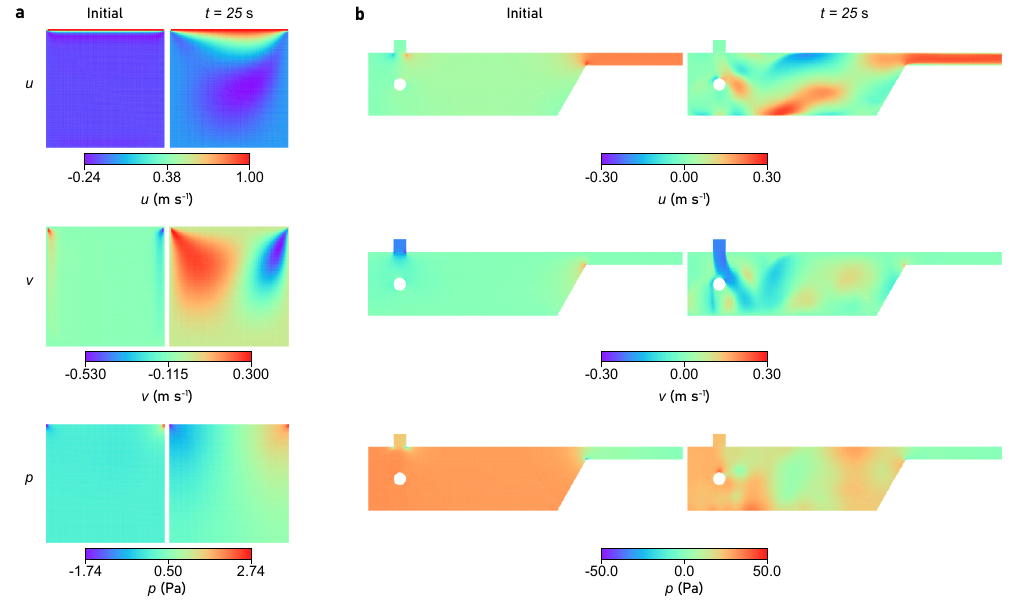}
   \caption{
   \textbf{Snapshots for numerical cases.}
   \textbf{a}, Snapshots of lid-driven cavity flow case (from top to bottom: x-component velocity $u$, y-component velocity $v$, and pressure $p$) at initial time and $t = 25$ s.
   \rev{\textbf{b}, Snapshots of complex cylinder flow case (from top to bottom: x-component velocity $u$, y-component velocity $v$, and pressure $p$) at initial time and $t = 25$ s.}
}
   \label{fig:1}
\end{figure}

\rev{
\begin{figure}[t!]
  \centering
   \includegraphics[width=0.8\linewidth]{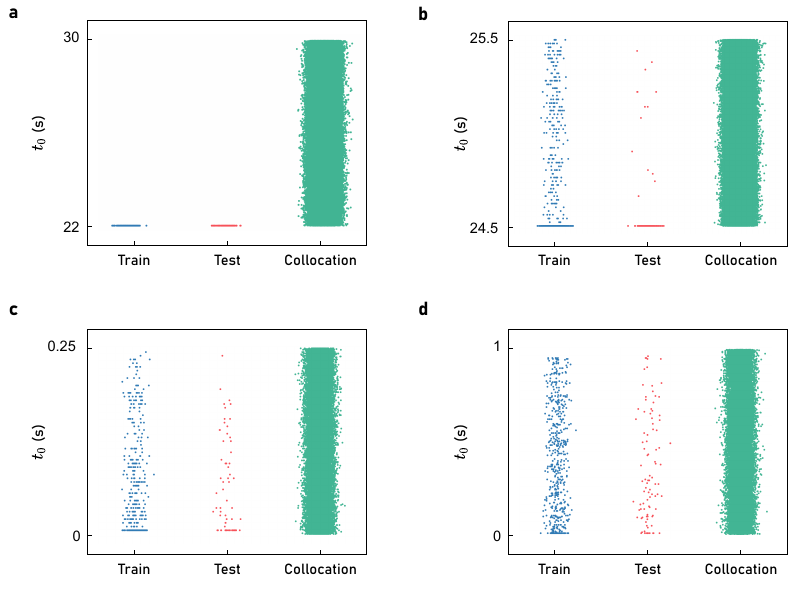}
   \caption{\rev{
   \textbf{\revdata{Distributions of particle release time $t_0$.}} Each dot denotes a particle release time. Horizontal axis shows the training set, test set, and collocation points (left to right). Vertical axis represent release time.
   \textbf{a}, Particle release time distribution in lid-driven cavity flow case.
   \textbf{b}, Particle release time distribution in complex cylinder flow case.
   \textbf{c}, Particle release time distribution in experimental aortic blood flow case.
   \textbf{d}, Particle release time distribution in experimental left ventricle blood flow case.
}}
   \label{fig:t0}
\end{figure}
}

\subsection{Lid-driven cavity flow}
The lid-driven cavity flow~\cite{botella1998benchmark} is a classic fluid dynamics problem that includes various phenomena such as boundary layer separation, cavity vortices, and self-sustained oscillations. We consider a 1 m $\times$ 1 m square cavity, where the left, right, and bottom walls were fixed, while the top wall acted as a moving boundary. The density of the working fluid is $1~\mathrm{kg/m^3}$ with the viscosity of $0.01~\mathrm{kg/(m\cdot s)}$. The top lid moved to the right at a velocity of $ u_0 = 1$m/s, resulting in a Reynolds number of 100, which was provided as a known condition to the network.
A snapshot of the flow field's velocity magnitude and pressure at  $t = 0$ s and $t = 25$ s is shown in  Figure \ref{fig:1}\textbf{a}. \rev{Figure \ref{fig:t0}\textbf{a} displays the distribution of particle release time $t_0$ in training set, test set and collocation points.}

\subsection{Complex cylinder flow}
Traditional studies typically focus on the wake region behind a single cylinder under simplified conditions. In contrast, our complex cylinder flow dataset simulates a more intricate flow field around a cylinder with open boundaries, employ water as the working fluid to better approximate practical engineering scenarios. 

The dataset was generated to simulate liquid water with a density of $998.2~\mathrm{kg/m^3}$ and a dynamic viscosity of $0.001003~\mathrm{kg/(m\cdot s)}$, which were provided to the network. A constant inflow velocity of $0.2~\mathrm{m/s}$ was imposed at the top inlet, with the outlet pressure set to $0$ Pa. All passive tracer particles were continuously released from the inlet to ensure an adequate particle number throughout the domain for training.
A snapshot of the flow field's velocity magnitude and pressure at  $t = 0$ s and $t = 25$ s  is shown in Figure~\ref{fig:1}\textbf{b}. \rev{Figure \ref{fig:t0}\textbf{b} illustrates how particle release times 
$t_0$ are distributed among the training set, test set, and collocation points.}

\subsection{Experimental aortic blood flow}
The aortic valve, situated between the left ventricle and the aorta, facilitates unidirectional blood flow into systemic circulation. Pathological calcification of the aortic valve leads to aortic stenosis (AS), a condition that affects $2.8$ \% of adults aged > 75 years, obstructing cardiac output and potentially causing fatal cardiovascular complications if left untreated. Transcatheter aortic valve replacement (TAVR) is a minimally invasive and worldwide adopted interventional treatment for aortic stenosis~\cite{takahashi2023differences,li2022comparison}. Developed as an alternative to surgical aortic valve replacement (SAVR) for high-risk patients, TAVR deploys bioprosthetic valves via catheter.  These valves are categorized into self-expanding valves (SEVs) with adaptive nitinol frames or balloon-expandable valves (BEVs), which require immediate circular expansion. While both valve types have been widely adopted, there is ongoing debate regarding their hemodynamic performance and long-term outcomes, with no definitive consensus on which valve type is superior. Our in vitro 3D particle image velocimetry study will quantify SEV/BEV flow fields, explaining mechanisms behind diagnostic discrepancies and guiding optimized valve selection.

A normal aortic root model without coronary connections was connected to a simulated circulatory loop driven by a pulsatile pump (Vivitro SuperPump, ViVitro Labs, Inc.). The silicone model with the left common carotid, left subclavian artery, and brachiocephalic artery was 3D-printed according to the CT data of a healthy person. Its refractive index is 1.41 and its thickness is $2$ mm. In the test section, the aortic root model was immersed into an acrylic hydraulic chamber filled with working fluid. The inlet of the model was connected with the left ventricular compliance chamber, and the model outlet was connected with the fluid reservoir. 

The blood analog was a mixture with $62$ \% water, $38$ \% glycerol by volume, and $230$ g/L urea by weight was added into the solution to match the refractive index. The density and dynamic viscosity of the mixture were $1.3284$ g/mL and $3.345 ~\mathrm{mPa\cdot s} $, respectively, and the refractive index was $1.4108$ at a working temperature of $26 ~^\circ C$. A 23 mm Edwards SAPIEN TAV was deployed at the annulus plane of the aortic root model. A clinician instructed the placement of the TAV. The valve position was secured using cable tie to prevent migration during pulsatile flow. A mechanical valve was used as a mitral valve within the ViVitro SuperPump. During systole, the blood analog flows from the left ventricular compliance chamber to the aortic root through the TAV, and then flow through the resistance valve, fluid reservoir and mitral valve back to left ventricle. Note that there is no blood flowing through the left common carotid, left subclavian artery and brachiocephalic artery. These vessels serve for fixation of the aortic root model during experiments. A resistance valve is used to adjust the systematic resistance. The fluid reservoir is open to the atmosphere to simulate the low pressure of the venous system. 

The flow downstream the TAV was measured under a physiological condition of 70 beats per minute and $7$ L/min cardiac output. The systolic and diastolic aortic pressure are $110$ mmHg and $45$ mmHg, respectively. A time-resolved two-dimensional (2D) PIV experiment was conducted to measure the flow downstream the TAV along the centerline of the root. The flow motion was visualized by seeding $7~\mathrm{\mu m}$ fluorescent Rhodamine-B-coated particles with a density of $1500 ~\mathrm{kg/m^3}$  into the working fluid. These particles were excited using a laser sheet with a thickness of $1.0$ mm, generated by a double-pulsed laser (Beamtech, Vlite-Hi-50) with a wavelength of $527 $ nm at a repetition rate of $0.2$ kHz. A high-speed CMOS camera (VEO-440L-36G-M version) equipped with a $100$ mm/f2.8 lens (Tokina atx-i) with a high-pass filter was used to capture the particle image pairs with resolution of $1664\times1660$ pixels. The acquisition frequency is $200$ Hz, and the time separation was set to $400~\mathrm{\mu s}$. The digital resolution of the images was $12.31$ pixels/mm, therefore, the maximum displacement was less than 15 pixels according to the maximum velocity of $3$ m/s at peak systole. Manual tracking of particle positions over time was also performed. 

\begin{figure}[t!]
  \centering
   \includegraphics[width=0.9\linewidth]{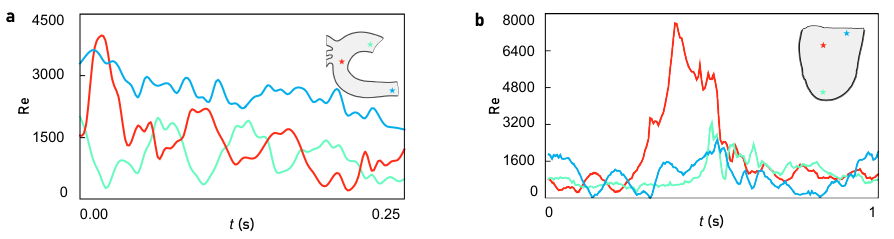}
   \caption{
   \textbf{Reynolds numbers measured upstream, downstream, and in the middle of the aorta, with line colors corresponding to position colors.}
   \textbf{a}, Reynolds numbers of experimental aortic blood flow dataset.
   \textbf{b}, Reynolds numbers of experimental left ventricle blood flow dataset.
}
   \label{fig:Re}
\end{figure}

The particle image pairs were processed using an in-house PIV software with a three-pass window deformation iterative multigrid scheme~\cite{scarano2001iterativeimage, wang2020globally}. Before velocity calculation, image preprocess with sliding minimum was performed to remove the image background. The final interrogation window size was $16\times16$ pixels with $50$ \% overlap, and the vectors at the final pass were Gaussian filtered with the same sizes the interrogation window to reduce the random noise. This configuration yielded a spatial resolution of $1.3$ mm with a vector interval of $0.65$ mm. The edge of aortic root was identified according to the image intensity gradient and converted to physical coordinates based on the calibration. The spatially varying Reynolds number obtained from the data, is shown in Figure \ref{fig:Re}\textbf{a}. Although our method is capable of predicting local Reynolds number values, we adopt a constant Reynolds number as an approximation, which introduces a source of model error. \rev{To construct the particle data, time-resolved PIV sequences containing tracers were processed using Fiji~\cite{schindelin2012fiji}, an open-source image analysis platform. Individual particle trajectories were manually tracked across consecutive frames through visual inspection, with spatial coordinates recorded at each time step. These pixel-level positions and temporal data were converted to physical units to generate particle trajectories and velocity fields. The generated particle trajectories and PIV velocimetry data were jointly used for model training and testing. The manual labeling of 465 spatiotemporal particle points took about 4.5 hours by one person. While more complex scenarios may require additional labeled trajectories for training, our method does not necessitate annotating every particle---only a finite set of labeled particle points are required for model training (e.g., a few hundred spatiotemporal points). Consequently, the annotation workload remains manageable, even under high particle density conditions.} \rev{The distribution of particles initial tracking times $t_0$ for the training set, test set, and collocation points is displayed in Figure \ref{fig:t0}\textbf{c}.}

\subsection{Experimental left ventricle blood flow}
The mitral valve, situated between the left atrium and ventricle, ensures unidirectional blood flow from the atrium to the ventricle~\cite{mccarthy2010anatomy}. In a normal mitral valve, intraventricular flow patterns exhibit physiological vortex dynamics that pump blood out of the left ventricle in an energy-efficient manner. Intraventricular hemodynamics is intricately linked to ventricular vortices and flow-induced stresses, which regulate the balance between hemodynamic load and myocardial stress~\cite{hove2003intracardiac}. This balance interfaces with biochemical pathways involving genes critical for cardiovascular morphogenesis, ultimately driving processes such as left ventricular remodeling. Consequently, analyzing left ventricular flow fields under physiological mitral valve conditions is essential for understanding their functional implications on cardiac performance.

For the experiment, the aortic root, aortic valve, and the complete mitral valve complex, including the annulus, anterior and posterior leaflets, chordae tendineae, and papillary muscles, were surgically dissected and sutured onto a 3D-printed annular resin plate connected to a silicone left ventricle. The geometry of the silicone left ventricle was reconstructed from adult heart CT data, with a wall thickness of $2$ mm and a refractive index of $1.4$.

The mitral valve-silicone left ventricle assembly was placed in a nine-sided chamber filled with a glycerol-water solution consisting of $62~\%$ water and $38~\%$ glycerol by volume. The solution had a dynamic viscosity of $3.49~\mathrm{mPa \cdot s}$ at a temperature of $26.9~^\circ C$. A piston driven by an electric motor generated periodic compression and relaxation of the chamber fluid to simulate left ventricular systole and diastole at a heart rate of $60$ beats per minute and a cardiac output of $3.6$ liters per minute. The fluid circulated in a closed loop, passing through the left ventricle, aortic valve, aortic compliance chamber, resistance valves, reservoir, left atrium, and mitral valve.

Fluorescent tracer particles coated with Rhodamine B, with an average diameter of $7$ microns, a density of $1500~\mathrm{kg/m^3}$, and a response frequency of $854.7$ Hz, were uniformly distributed in the flow field. A double-pulsed laser system, model Vlite-Hi-527-50 from Beijing Laser Optoelectronics Technology Co., Ltd., with a wavelength of $527$ nm, a repetition rate of $200$ Hz, and a pulse interval optimized to ensure displacement less than $8$ pixels, generated a light sheet with a thickness of $1.0$ mm. The emitted fluorescence at $590$ nm was captured by a high-speed camera, model Imager HS from LaVision GmbH, equipped with a Tokina $100$ mm/f2.8 fixed-focus lens and an optical filter with transmission greater than $90~\%$ for wavelengths above $570$ nm. The image resolution was $2016\times2016$ pixels, with $20$ pixels per millimeter. Velocity fields were calculated using a window deformation iterative multigrid scheme, with two iterations and a final interrogation window size of $32\times32$ pixels overlapped by $75~\%$. Outliers were removed using normalized median tests, achieving a vector spacing of $0.4$ mm and a spatial resolution of $1.6$ mm. Gaussian smoothing was applied to enhance the visualization of flow structures, while particle trajectories were manually tracked \rev{using Fiji, as in the previous experimental case. One person spent approximately 4.5 hours manually labeling 659 spatiotemporal particle points.} \rev{Figure \ref{fig:t0}\textbf{a} shows the distribution of particle tracking start times $t_0$ across the training set, test set, and collocation points.}
The spatially varying Reynolds number obtained from the data, is shown in Figure \ref{fig:Re}\textbf{b}.

\section{Results}
Our model utilizes the Limited-memory Broyden-Fletcher-Goldfarb-Shanno (LBFGS) optimizer~\cite{liu1989limited} with consistent parameter settings across all models. The optimizer is configured with a learning rate of $1.0$, maximum iterations set to $500,000$, maximum function evaluations limited to $500,000$, a history size of $50$, tolerance for gradient convergence at $1\times 10^{-5}$, tolerance for change in function value at $1.0$ times the machine epsilon, and employs the strong Wolfe line search method. \revdata{Detailed network configuration parameters and training details are provided in our publicly released code.}


\subsection{Fourier feature mapping}
The method of Fourier feature mapping (FFM) was employed in our article. Two feature mapping matrices $\mathbf{B}_1\sim \mathcal{N}(0,\sigma_{ffm_1}^2)$ and $\mathbf{B}_2\sim \mathcal{N}(0,\sigma_{ffm_2}^2)$ were initialized with Gaussian-distributed values to modulate the frequency spectrum of encoded inputs, where $\sigma_{ffm_1}$ and $\sigma_{ffm_2}$ control the frequency ranges of the respective mappings. The dimensionality and variance of $\mathbf{B}_1$ and $\mathbf{B}_2$ were empirically tuned for each dataset to balance representation capacity and computational efficiency, parameters chosen details were shown in Table~\ref{FFNPara}.
These configurations were selected through sensitivity analysis on validation datasets.

\begin{table}[htpb]
\centering
\caption{Fourier feature mapping parameters for different flow datasets.}
\begin{tabular}{lcccc}
\toprule
Verifiable Example & $\mathbf{B}_1$ Size & $\sigma_{ffm_1}$ & $\mathbf{B}_2$ Size & $\sigma_{ffm_2}$ \\
\midrule
Lid-driven cavity flow & $50 \times 4$ & 0.1 & $50 \times 3$ & 0.1 \\
Complex cylinder flow & $5000 \times 4$ & 0.1 & $500 \times 3$ & 0.1 \\
Experimental aortic blood flow & $500 \times 4$ & 0.1 & $500 \times 3$ & 0.1 \\
Experimental left ventricle blood flow & $1000 \times 4$ & 1.0 & $1000 \times 3$ & 1.0 \\
\bottomrule
\label{FFNPara}
\end{tabular}
\end{table}

We compared the impact of using Fourier features on the prediction results in test set, as shown in Table~\ref{FFN}. The results indicate that using this method maintains good prediction accuracy in most of the cases.

\begin{table}[t!]
\centering
\caption{Comparison of prediction accuracy with and without Fourier feature mapping in test set.}
\label{FFN}
\small
\begin{tabular}{@{}lcccccc@{}}
\toprule
\multirow{2}{*}{Verifiable Example} & \multirow{2}{*}{Method} & \multicolumn{4}{c}{Relative  $\ell_2$ Error} \\
\cmidrule(l){3-6}
& & $x$ & $y$ & $u$ & $v$ \\
\midrule
\multirow{2}{*}{\makecell[l]{Lid-driven\\cavity flow}} 
& Without FFM  & \pmb{$2.28 \times 10^{-2}$} & {$1.57 \times 10^{-2}$}  & $6.26 \times 10^{-2}$ & $5.83 \times 10^{-2}$ \\
& With FFM  & {$2.42 \times 10^{-2}$} & \pmb{$1.54 \times 10^{-2}$}  & \pmb{$1.85 \times 10^{-2}$} & \pmb{$1.87 \times 10^{-2}$} \\
\midrule
\multirow{2}{*}{\makecell[l]{Complex \\ cylinder field}}
& Without FFM   & {\pmb{$2.93 \times 10^{-2}$} }& {\pmb{$4.96 \times 10^{-2}$}  }& {\pmb{$1.21 \times 10^{-1}$} }& {\pmb{$1.79 \times 10^{-1}$}} \\
& With FFM  & {$3.27 \times 10^{-2}$ }& {$8.01 \times 10^{-2}$  }& {{$1.66 \times 10^{-1}$} }& {{$2.05 \times 10^{-1}$}} \\
\midrule
\multirow{2}{*}{\makecell[l]{Experimental \\ aortic blood flow}}
& Without FFM   & $6.65 \times 10^{-2}$ & $3.91 \times 10^{-2}$  & $5.94 \times 10^{-1}$ & $8.53 \times 10^{-1}$ \\
& With FFM  & \pmb{$4.54 \times 10^{-2}$} & \pmb{$2.52 \times 10^{-2}$}  & \pmb{$4.41 \times 10^{-1}$} & \pmb{$5.93 \times 10^{-1}$} \\
\midrule
\multirow{2}{*}{\makecell[l]{Experimental \\  LV blood flow}}
& Without FFM   & $2.08 \times 10^{-2}$ & {$3.57 \times 10^{-2}$}  & $8.03 \times 10^{-1}$ & $4.84 \times 10^{-1}$ \\
& With FFM  & \pmb{$1.47 \times 10^{-2}$} & \pmb{$3.08 \times 10^{-2}$}  & \pmb{$7.38 \times 10^{-1}$} & \pmb{$3.93 \times 10^{-1}$} \\
\bottomrule
\end{tabular}
\end{table}

\subsection{Comparison of separate and joint training strategies}
In this section, we compare the TrajectoryFlowNet with separate training the Trajectory and Flow Field blocks. For the separate prediction methods, we ensure that there is no connection between network Trajectory block and Flow Field block, with the Trajectory block takes $(\mathbf{x}_0, t_0, \tau)$ as inputs, while the Flow Field block operates on ground-truth spatial-temporal coordinates $(\mathbf{x},t)$. Both blocks maintain the same output configurations as in the algorithm described in the paper. Meanwhile, the separately trained Trajectory block does not have physical constraints, whereas Flow Field block includes NS equation constraints. All other settings remain consistent with the algorithm in the paper. The results for test set are shown in Table~\ref{Separate training}. \revdata{Our algorithm generally outperforms separate prediction methods on the test set.}


\begin{table}[htpb]
\centering
\caption{Comparison of performance between TrajectoryFlowNet and separate training methods in test set.}
\label{Separate training}
\small
\begin{tabular}{@{}lcccccc@{}}
\toprule
\multirow{2}{*}{Verifiable Example} & \multirow{2}{*}{Method} & \multicolumn{4}{c}{Relative  $\ell_2$ Error} \\
\cmidrule(l){3-6}
& & $x$ & $y$ & $u$ & $v$ \\
\midrule
\multirow{2}{*}{\makecell[l]{Lid-driven\\cavity flow}} 
& Separate training  & \pmb{$1.56 \times 10^{-2}$} & \pmb{$1.07 \times 10^{-2}$}  & $1.93 \times 10^{-2}$ & $2.17 \times 10^{-2}$ \\
& TrajectoryFlowNet  & $2.42 \times 10^{-2}$ & $1.54 \times 10^{-2}$  & \pmb{{$1.85 \times 10^{-2}$}} & \pmb{$1.87 \times 10^{-2}$} \\
\midrule
\multirow{2}{*}{\makecell[l]{Complex \\ cylinder flow}}
& Separate training   & \pmb{$3.02 \times 10^{-2}$  }& \pmb{$7.76 \times 10^{-2}$}& {\pmb{$1.09 \times 10^{-1}$} }& {\pmb{$2.00 \times 10^{-1}$}} \\
& TrajectoryFlowNet  & $3.27 \times 10^{-2}$  & $8.01 \times 10^{-2}$& {$1.66 \times 10^{-1}$} & {$2.05 \times 10^{-1}$} \\
\midrule
\multirow{2}{*}{\makecell[l]{Experimental \\ aortic blood flow}}
& Separate training   &$4.78 \times 10^{-2}$  &$3.23 \times 10^{-2}$  & \pmb{$4.34 \times 10^{-1}$} & $6.10 \times 10^{-1}$ \\
& TrajectoryFlowNet  & \pmb{$4.54 \times 10^{-2}$} & \pmb{$2.52 \times 10^{-2}$}  & {$4.41 \times 10^{-1}$} & \pmb{$5.93 \times 10^{-1}$} \\
\midrule
\multirow{2}{*}{\makecell[l]{Experimental \\  LV blood flow}}
& Separate training   & $1.69 \times 10^{-2}$ & $ 3.34 \times 10^{-2}$  & $8.66 \times 10^{-1}$ & $4.16 \times 10^{-1}$ \\
& TrajectoryFlowNet & \pmb{$1.47 \times 10^{-2}$} & \pmb{$3.08 \times 10^{-2}$}  & \pmb{$7.38 \times 10^{-1}$} & \pmb{$3.93 \times 10^{-1}$}  \\
\bottomrule
\end{tabular}
\end{table}

\subsection{Activation function}

\begin{table}[htbp]
\centering
\caption{Evaluation of activation functions based on relative  $\ell_2$ error in test set.}
\label{Activation Functions}
\small
\begin{tabular}{@{}lcccccc@{}}
\toprule
\multirow{2}{*}{Verifiable Example} & \multirow{2}{*}{Activation Function} & \multicolumn{4}{c}{Relative  $\ell_2$ Error} \\
\cmidrule(l){3-6}
& & $x$ & $y$ & $u$ & $v$ \\
\midrule
\multirow{4}{*}{\makecell[l]{Lid-driven\\cavity flow}} 
& ReLU  & $1.33 \times 10^{-1}$ & $1.06 \times 10^{-1}$  & $6.13 \times 10^{-1}$ & $7.13 \times 10^{-1}$ \\
& ELU   & $6.17 \times 10^{-2}$ & $5.36 \times 10^{-2}$  & $2.83 \times 10^{-1}$ & $2.51 \times 10^{-1}$ \\
& Tanh  & \pmb{$2.42 \times 10^{-2}$} & \pmb{$1.54 \times 10^{-2}$}  & \pmb{$1.85 \times 10^{-2}$} & \pmb{$1.87 \times 10^{-2}$}\\
& Sigmoid  & $5.45 \times 10^{-1}$ & $4.91 \times 10^{-1}$  & $9.99 \times 10^{-1}$ & $9.99 \times 10^{-1}$ \\
\midrule
\multirow{4}{*}{\makecell[l]{Complex \\ cylinder flow}}
& ReLU  & {$4.43 \times 10^{-2}$ }&{ $1.14 \times 10^{-1}$  }& {$4.32 \times 10^{-1}$ }&{ $5.54 \times 10^{-1}$ }\\
& ELU   & \pmb{$2.59 \times 10^{-2}$ }&\pmb{ $7.33 \times 10^{-2}$  }& {$2.69 \times 10^{-1}$ }& {$4.39 \times 10^{-1}$ }\\
& Tanh  & {$3.27 \times 10^{-2}$ }& {$8.01 \times 10^{-2}$  }& \pmb{{$1.66 \times 10^{-1}$} }& \pmb{{$2.05 \times 10^{-1}$}} \\
& Sigmoid  &{ $1.73 \times 10^{-1}$ }&{ $4.09 \times 10^{-1}$ } &{ $1.00 \times 10^{0}$ }&{ $1.00 \times 10^{0}$ }\\
\midrule
\multirow{4}{*}{\makecell[l]{Experimental \\ aortic blood flow}}
& ReLU  & $4.75 \times 10^{-2}$ & $ 3.03 \times 10^{-2}$  & $6.86 \times 10^{-1}$ & $7.43 \times 10^{-1}$ \\
& ELU   & \pmb{$4.46 \times 10^{-2}$} & $2.56 \times 10^{-2}$  & $4.92 \times 10^{-1}$ & $6.69 \times 10^{-1}$ \\
& Tanh  & {$4.54 \times 10^{-2}$} & \pmb{$2.52 \times 10^{-2}$}  & \pmb{$4.41 \times 10^{-1}$} & \pmb{$5.93 \times 10^{-1}$}  \\
& Sigmoid  & $7.66 \times 10^{-2}$ & $3.08 \times 10^{-2}$  & $1.00 \times 10^{0}$ & $1.00 \times 10^{0}$ \\
\midrule
\multirow{4}{*}{\makecell[l]{Experimental \\ LV blood flow}}
& ReLU  & $2.61 \times 10^{-2}$ & {$5.48 \times 10^{-2}$}  & $9.99 \times 10^{-1}$ & $9.99 \times 10^{-1}$ \\
& ELU   & \pmb{$1.42 \times 10^{-2}$} & $3.23 \times 10^{-2}$  & \pmb{$6.98 \times 10^{-1}$} & $3.95 \times 10^{-1}$ \\
& Tanh  & {$1.47 \times 10^{-2}$} & \pmb{$3.08 \times 10^{-2}$}  & {$7.38 \times 10^{-1}$} & \pmb{$3.93 \times 10^{-1}$}  \\
& Sigmoid  & $2.64 \times 10^{-2}$ & $5.50 \times 10^{-2}$  & $1.00 \times 10^{0}$ & $1.00 \times 10^{0}$ \\
\bottomrule

\end{tabular}
\end{table}

In this section, we conducted tests on various activation functions, including the Rectified Linear Unit (ReLU)~\cite{hara2015analysis}, Exponential Linear Unit (ELU)~\cite{clevert2015fast}, Hyperbolic Tangent (Tanh)~\cite{nwankpa2018activation}, and Sigmoid~\cite{han1995influence}. The effectiveness of these activation functions was evaluated based on the total sum of relative  $\ell_2$ error from the test set, as shown in Table~\ref{Activation Functions}. It is worth noting that the only difference between the different test models lies in the choice of activation function, as all models were trained using the L-BFGS optimizer until convergence. The experimental results consistently demonstrated that the Tanh activation function outperformed others across all datasets. Therefore, we consider Tanh to be a suitable and widely applicable choice for the majority of datasets in this study.

\rev{
\subsection{Impact of particle sparsity levels}

In this section, we evaluate the TrajectoryFlowNet with different particle sparsity levels. We selected the lid-driven cavity flow case as a benchmark case because the total number of particles remains constant (no inflow or outflow) and all particles experience the same motion duration, thereby eliminating effects from varying trajectory lengths. Additionally, as a simulated flow, it provides access to abundant high-fidelity data for training and validation.

Starting from an initial training set comprising 200 particle trajectories, we systematically reduced the number of training samples to 80\%, 50\%, 30\%, and 10\% of the original size while holding the test set constant, thereby quantifying the impact of particle sparsity on model performance.

Figure~\ref{fig:Sparse}\textbf{a} shows the training particle positions at $25$s with different sparsity levels. Figure~\ref{fig:Sparse}\textbf{b} represents the predicted particle trajectories in the same test set, while Figure~\ref{fig:Sparse}\textbf{c} and Figure~\ref{fig:Sparse}\textbf{d} illustrates the predicted flow velocity and corresponding error at $25$s, respectively. 
\revdata{Our model maintains high predictive accuracy even with the reduced training dataset.}
However, particle trajectory prediction is more sensitive to data sparsity than velocity field prediction. This is expected that trajectory prediction requires long-term, precise tracking and is inherently more challenging than predicting velocity fields. Furthermore, the flow velocity fields are predicted by the model for a given query time point independently, whereas the trajectory prediction accumulates errors over a certain time duration, making trajectory errors appear more pronounced. The relative  $\ell_2$ errors of the predictions are summarized in Table~\ref{TSparse}.

\begin{figure}[t!]
  \centering
   \includegraphics[width=1.0\linewidth]{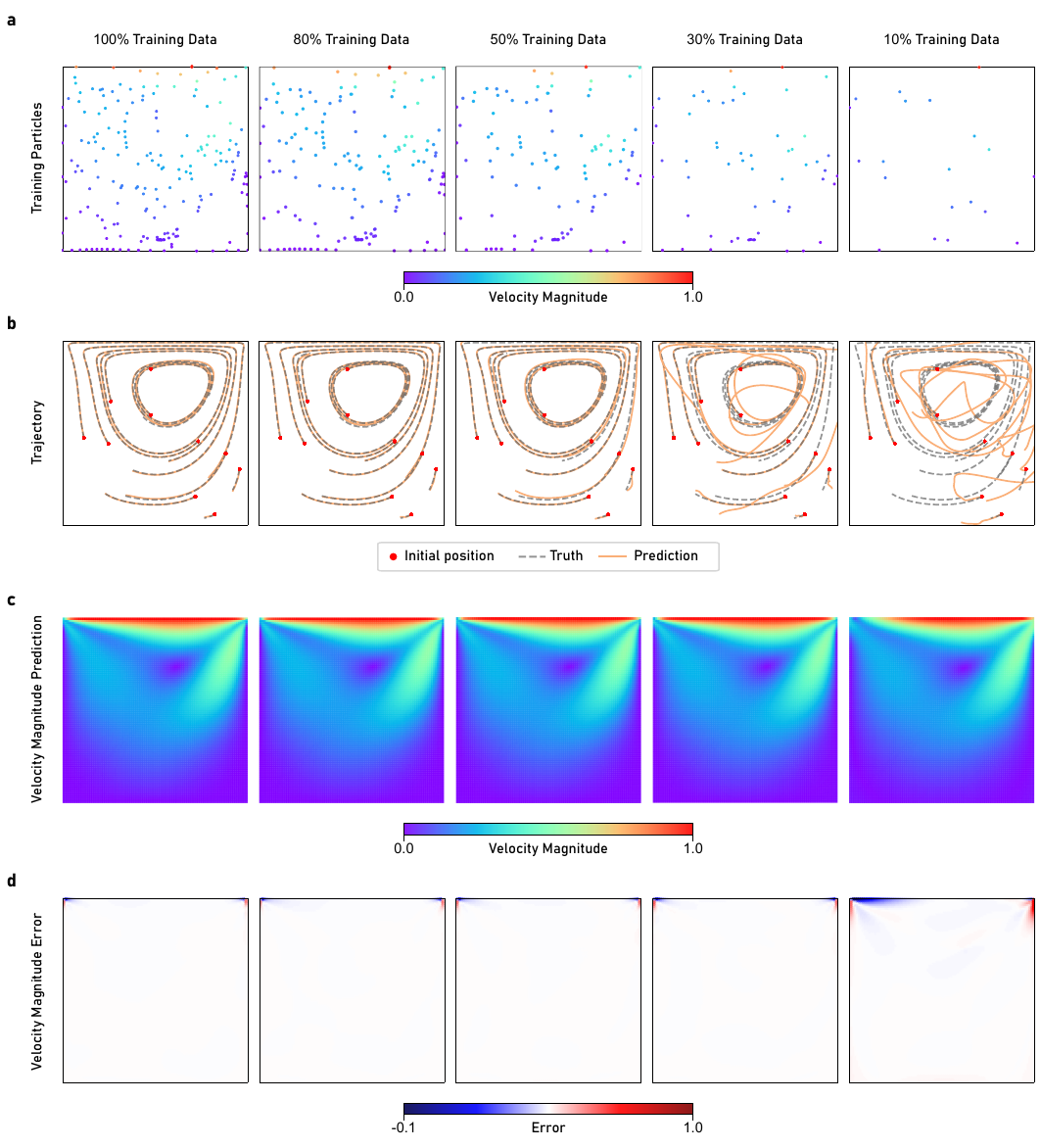}
   \caption{
   \revdata{\textbf{Results of the lid-driven cavity flow case with different particle sparsity.}}
   (From left to right: 100\% training data, 80\% training data, 50\% training data, 30\% training data and 10\% training data.)
   \textbf{a}, Particle positions at $t = 25$s of corresponding sparsity level. Color represents the particle velocity magnitude.
   \textbf{b}, Predicted particle trajectories in the same test set. Red dots denote initial particle
 positions, and gray dots the trajectory points.
   \textbf{c}, Snapshots of the flow velocity magnitude prediction at $t = 25$s.
   \textbf{d}, Prediction error of the flow velocity magnitude prediction at $t = 25$s.
}
   \label{fig:Sparse}
\end{figure}

\begin{table}[t!]
\centering
\caption{\rev{Comparison of prediction accuracy with different particle sparsity levels.}}
\label{TSparse}
\begin{tabular}{@{}lccccc@{}}
\toprule
\multirow{2}{*}{Sparsity Levels}  & \multicolumn{4}{c}{Relative  $\ell_2$ Error} \\
\cmidrule(l){2-5}
& $x$ & $y$ & $u$ & $v$ \\
\midrule
100\%  & \pmb{$2.42 \times 10^{-2}$} & \pmb{$1.54 \times 10^{-2}$}  & \pmb{$1.85 \times 10^{-2}$} & \pmb{$1.87 \times 10^{-2}$} \\
80\%  & $3.36 \times 10^{-2}$ & $2.45 \times 10^{-2}$  & $2.23 \times 10^{-2}$ & $2.11 \times 10^{-2}$ \\
50\%  & $1.02 \times 10^{-1}$ & $6.45 \times 10^{-2}$  & $2.39 \times 10^{-2}$ & $2.17 \times 10^{-2}$ \\
30\%  & $2.35 \times 10^{-1}$ & $1.75 \times 10^{-1}$  & $2.46 \times 10^{-2}$ & $2.34 \times 10^{-2}$ \\
10\%  & $5.99 \times 10^{-1}$ & $3.91 \times 10^{-1}$  & $1.09 \times 10^{-1}$ & $8.24 \times 10^{-2}$ \\

\bottomrule
\end{tabular}
\end{table}

}

\clearpage

